\documentclass[12pt]{article}

\usepackage{latexsym}

\usepackage[font=small,labelfont=bf]{caption}

\usepackage{graphics}
\usepackage{graphicx}
\usepackage{epstopdf}
\usepackage{amssymb}
\usepackage{tabularx}
\usepackage{caption}
\usepackage{subcaption}
\usepackage{dcolumn}
\usepackage{bm}
\usepackage{hyperref}
\usepackage{tabularx}
\usepackage{slashbox}
\usepackage{tikz}
\usetikzlibrary{matrix}
\newsavebox{\tempbox}
\usepackage{gensymb}
\newcommand{\be}{\begin{equation}}
\newcommand{\ee}{\end{equation}}
\newcommand{\ba}{\begin{eqnarray}}
\newcommand{\ea}{\end{eqnarray}}
\newcommand{\ban}{\begin{eqnarray*}}
\newcommand{\ean}{\end{eqnarray*}}

\graphicspath{{./Figures/}}
\begin{document}


\title {Observational signatures of strongly naked singularities: image of the thin accretion disk}

\author{
Galin Gyulchev$^{1}$\footnote{E-mail: \texttt{gyulchev@phys.uni-sofia.bg}}, \, Jutta Kunz$^{2}$\footnote{E-mail: \texttt{jutta.kunz@uni-oldenburg.de}}, \, Petya Nedkova$^{1,2}$\footnote{E-mail: \texttt{pnedkova@phys.uni-sofia.bg}},  \, Tsvetan Vetsov$^{1}$\footnote{E-mail: \texttt{vetsov@phys.uni-sofia.bg}},\\ Stoytcho Yazadjiev$^{1,3}$\footnote{E-mail: \texttt{yazad@phys.uni-sofia.bg}}\\ \\
   {\footnotesize${}^{1}$ Faculty of Physics, Sofia University,}\\
  {\footnotesize    5 James Bourchier Boulevard, Sofia~1164, Bulgaria }\\
  {\footnotesize${}^{2}$Institute of Physics, Carl von Ossietzky University of Oldenburg,} \\
  {\footnotesize 26111 Oldenburg, Germany}\\
  {\footnotesize${}^{3}$ Institute of Mathematics and Informatics,}\\
{\footnotesize Bulgarian Academy of Sciences, Acad. G. Bonchev 8, } \\
  {\footnotesize  Sofia 1113, Bulgaria}}
\date{}
\maketitle

\begin{abstract}
We study the optical appearance of a thin accretion disk around the strongly naked static Janis-Newman-Winicour singularity. The solution does not possess a photon sphere, which results in the formation of a complex structure of bright rings in the central region of the disk image. Such structure is absent in the case of the Schwarzschild black hole with a thin accretion disk, where instead of the image we observe the black hole shadow. Some of the rings emit with the maximal observable radiation flux from the accretion disk, and should be experimentally detectable. Thus, this qualitatively new feature can be used to distinguish observationally black holes from naked singularities. We elucidate the appearance of the ring structure by revealing the physical mechanism of its formation, and explaining the nature of each of the ring images. We make the conjecture that a similar structure would also appear for other solutions without a photon sphere and it can serve as a general observational signature for distinguishing compact objects possessing no photon sphere from  black holes.
\end{abstract}

\section{Introduction}

Black holes are one of the most amazing predictions of general relativity, which was confirmed by numerous independent experimental efforts, including the recently detected  gravitational waves and the observation of the shadow images \cite{EHT01}-\cite{EHT6}. However, gravitational theories predict a variety of other compact objects which can possess a similar phenomenological behavior. Such more exotic compact objects can involve for example wormholes, naked singularities or gravastars. They are motivated by the development of a quantum gravity theory, which is supposed to resolve curvature singularities, and provide physical mechanisms for the formation of matter fields with negative energy density.

Exotic compact objects don't possess an event horizon. However, other characteristic surfaces and distances, which are typical for black holes, can be present in their spacetimes. An example of such a surface is the photon sphere. It arises in relation to the geodesic motion in the background of a compact object, and serves as a separatrix between infalling geodesic bundles which scatter away to infinity, and those which are captured by the compact object. In addition, it was demonstrated that the photon sphere can be used to distinguish between different static compact objects' spacetimes. It can replace the event horizon in the formulation of the corresponding uniqueness theorems, and in this way it enables a much broader classification including also horizonless objects apart from black holes \cite{Cederbaum:2014}-\cite{Yazadjiev:2016}.

The configuration of the photon sphere, or its generalization to a photon region when non-spherically symmetric spacetimes are considered, determines the qualitative behavior of the light propagation in the background of the compact object. Thus, compact objects with similar photon regions give rise to qualitatively similar shadow images. Such behavior is observed in the case of black holes in different alterative theories of gravity, where, irrespective of the theory under consideration, similarities in the photon region structure lead to almost indistinguishable observable images \cite{Amarilla:2010}-\cite{Ahmedov:2014}. In the same way, exotic compact objects such as naked singularities and wormholes can closely resemble black holes in their shadow silhouettes when their photon regions are alike \cite{Nedkova:2013}-\cite{Shaikh:2019}. On the other hand, systems of interacting compact objects such as two compact objects in equilibrium \cite{Yumoto}-\cite{Shipley}, merging black holes \cite{Bohn}, black holes with a scalar field condensate \cite{Cunha:2015}-\cite{Cunha:2016}, or quasi-stationary configurations of a black hole and an accretion disk when the back reaction of the accreting matter is taken into account \cite{Grover:2018}, can produce observable shadow images with drastically different structure. The reason is that in such cases the geodesics motion is not integrable and chaotic scattering can occur. The photon region is not simply connected and leads to the formation of multiple disconnected shadow images which can possess a fractal structure.

In realistic astrophysical situations compact objects are not supposed to exist in isolation but surrounded by a disk of accreting matter. The accretion disk modifies the optical appearance of the compact object. For example, a thin accretion disk serves as a geometrically thin but optically thick medium, and in this case the compact object shadow corresponds to the observable image of the innermost stable circular particle orbit (ISCO), and not of the photon sphere\footnote{We adopt an astrophysical definition of the shadow as the dark region observed in the compact object's image.}. The optical appearance of the accretion disk adds further information about the observable properties of the spacetime, and its distinctive features can be used for distinguishing more precisely between black holes and more exotic compact objects by electromagnetic observations. An interesting question is whether there exists some mathematical structure which controls the similarities between the observable images. The structure of the photon region or its mere existence determines again the formation of the shadow image, although the shadow, when it is observed, is practically a visualization of the ISCO.

In this paper we study the optical appearance of a thin accretion disk around a static spherically symmetric naked singularity. The naked singularity is described by an exact solution of the Einstein-scalar field equations, which was originally obtained by Fisher in  \cite{Fisher:1948}, but it was later rediscovered in various works, e.g. \cite{Janis:1968}- \cite{Virbhadra:1997}, becoming most famous in the literature in the form derived by Janis, Newman and Winicour.  The optical properties of the Janis-Newman-Winicour (JNW) solution were investigated in detail in a series of works \cite{Virbhadra:1998}-\cite{Ovgun:2008}, including the gravitational lensing and the relativistic images that can arise in its spacetime. The solution is characterized by two physically distinct regimes depending on its scalar charge to mass ratio called weakly and strongly naked singularities. The weakly naked singularity possesses a photon sphere, and its observable properties in the electromagnetic spectrum are expected to resemble those of the Schwarzschild black hole. Indeed, it was shown in \cite{Shaikh:2019} that if the compact object is surrounded by a geometrically thick and optically thin medium, its shadow closely approaches the image of the Schwarzschild black hole, while in our previous work \cite{Nedkova:2019} we investigated the optical appearance of a thin accretion disk in its vicinity. Our analysis demonstrated that the accretion disk will appear in a qualitatively similar way to a distant observer and only quantitative differences will be present in its size and radiation intensity. The problem was also revisited in \cite{Shaikh:2019b}, and the optical appearance of wormholes with a thin accretion disk was investigated in \cite{Shaikh:2019c}.  This work is a natural continuation of our previous studies examining the observable features of a thin disk around the strongly naked Janis-Newman-Winicour singularity. We will show that the absence of a photon sphere in the strongly naked case leads to important distinctions. We don't observe the formation of a shadow, i.e. a dark region in the interior of the ISCO image, since a structure of bright rings appears at the center of the accretion disk apparent shape. We further explain the explicit physical mechanism of the formation of the central rings image, and demonstrate that in our case it is directly connected with the absence of a photon sphere. However, we should note that this is not the only mechanism, which can lead to the appearance of ring images. They can be observed for other exotic compact objects such as boson stars and wormholes possessing a photon region \cite{Shaikh:2019c}, \cite{Vincent:2020}, where they are formed by different physical reasons.  The images are obtained applying the visualization techniques developed in \cite{Nedkova:2019}, which build on earlier studies on the optical appearance of a thin accretion disk around the Schwarzschild black hole \cite{Cunningham:1972}-\cite{Muller:2012}.

The paper is organized as follows. In the next section we briefly present the Janis-Newman-Winicour solution focusing on the properties of the circular geodesics in the equatorial plane, which are necessary for the application of the thin accretion disk model. In section 3 we study the optical appearance of the thin accretion disk around a strongly naked Janis-Newman-Winicour singularity. We describe our techniques for constructing the images, and present as a case study the appearance of the strongly naked singularity with a thin accretion disk for the solution parameter $\gamma = 0.48$, which is representative for a certain region of the parametric space. We observe a variety of images of different nature and origin. In section 4 we analyze the morphology of the disk image, elucidating the physical mechanism for its formation and the processes giving rise to its features. The most characteristic feature is the appearance of a set of bright rings in the central part of the image instead of a shadow, which are formed due to the absence of a photon sphere. In section 5 we study the Novikov-Thorne model of radiation of the thin disk, and evaluate the intensity of the observable flux in the different parts of the disk image. In section 6 we summarize our results and give our conclusions.

\section{Circular orbits in the Janis-Newman-Winicour spacetime}

The Janis-Newman-Winicour naked singularity  \cite{Janis:1968} is  the unique  spherically symmetric solution to the Einstein-massless scalar field equations

\begin{eqnarray}
&&{\cal R}_{\mu\nu} = 2\nabla_{\mu}\varphi\nabla_{\nu}\varphi, \\
&&\nabla_{\mu}\nabla^{\mu}\varphi =0. \nonumber
\end{eqnarray}
The theory allows no spherically symmetric black holes except for the Schwarzschild black hole, which is recovered in the limit of a vanishing scalar field. The metric of the Janis-Newman-Winicour solution can be represented in the parametrization \cite{Virbhadra:1997}

\begin{equation}
ds^2 = -\left(1-\frac{2M}{\gamma r}\right)^{\gamma} dt^2
      +\left(1-\frac{2M}{\gamma r}\right)^{-\gamma} dr^2
      + \left(1-\frac{2M}{\gamma r}\right)^{1-\gamma} r^2\left(d\theta^2 + \sin^2\theta d\phi^2\right),
\end{equation}
where the scalar field takes the form
\begin{equation}
\varphi = \frac{q\gamma}{2M} \ln\left(1-\frac{2M}{\gamma r}\right).
\end{equation}
The solution is characterized by two real parameters $M$ and $q$, where $M$ represents its $ADM$ mass and $q$ is its scalar charge. The parameter $\gamma$ is related to  the scalar charge to mass ratio according to
\begin{eqnarray}\label{param}
\gamma = \frac{M}{\sqrt{M^2+q^2}},
\end{eqnarray}
taking the range $0\leq\gamma\leq1$. In the limit $\gamma=1$ we obtain the Schwarzschild solution, which corresponds to a zero scalar charge.
For non-trivial scalar fields the solution describes a naked curvature singularity located at the radial coordinate $r_{cs}=2M/\gamma$.

The Janis-Newman-Winicour solution possesses qualitatively different physical properties depending on its scalar charge to mass ratio. For low values of the scalar charge satisfying $0<q/M < \sqrt{3}$ the solution describes a weakly naked singularity \cite{Virbhadra:2002}, while when the scalar charge starts to dominate as  $q/M > \sqrt{3}$, the solution gets into the regime of a strongly naked singularity.  The weakly naked singularities resemble the Schwarzschild black hole in their lensing properties. They possess a photon sphere, and a similar structure of the relativistic images like the Einstein rings for example. In the strongly naked regime the photon sphere disappears and the light propagation in their vicinity leads to qualitatively different effects. The border case $q/M = \sqrt{3}$, called a marginally strongly naked singularity, behaves similarly as the weakly naked scenario. Using the relations between the solution parameters ($\ref{param}$), we can express the conditions for the different physical regimes in terms of the parameter $\gamma$, which takes the range $0.5< \gamma< 1$ for weakly naked singularities, and $0\leq \gamma< 0.5$ for the strongly naked case.

\begin{figure}[h!]
    		\setlength{\tabcolsep}{ 0 pt }{\footnotesize\tt
		\begin{tabular}{ ccc }
           \includegraphics[width=0.32\textwidth]{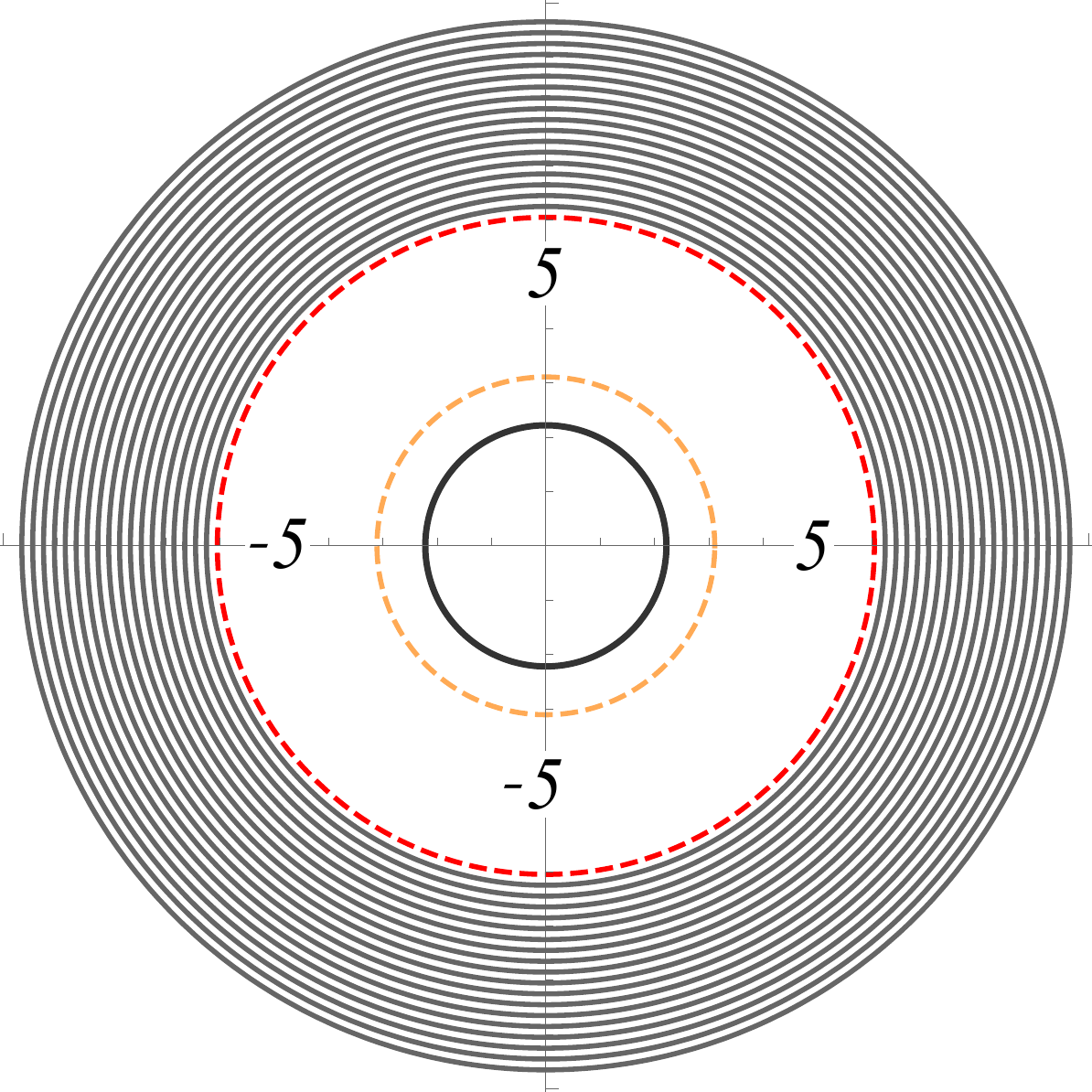}
             \includegraphics[width=0.32\textwidth]{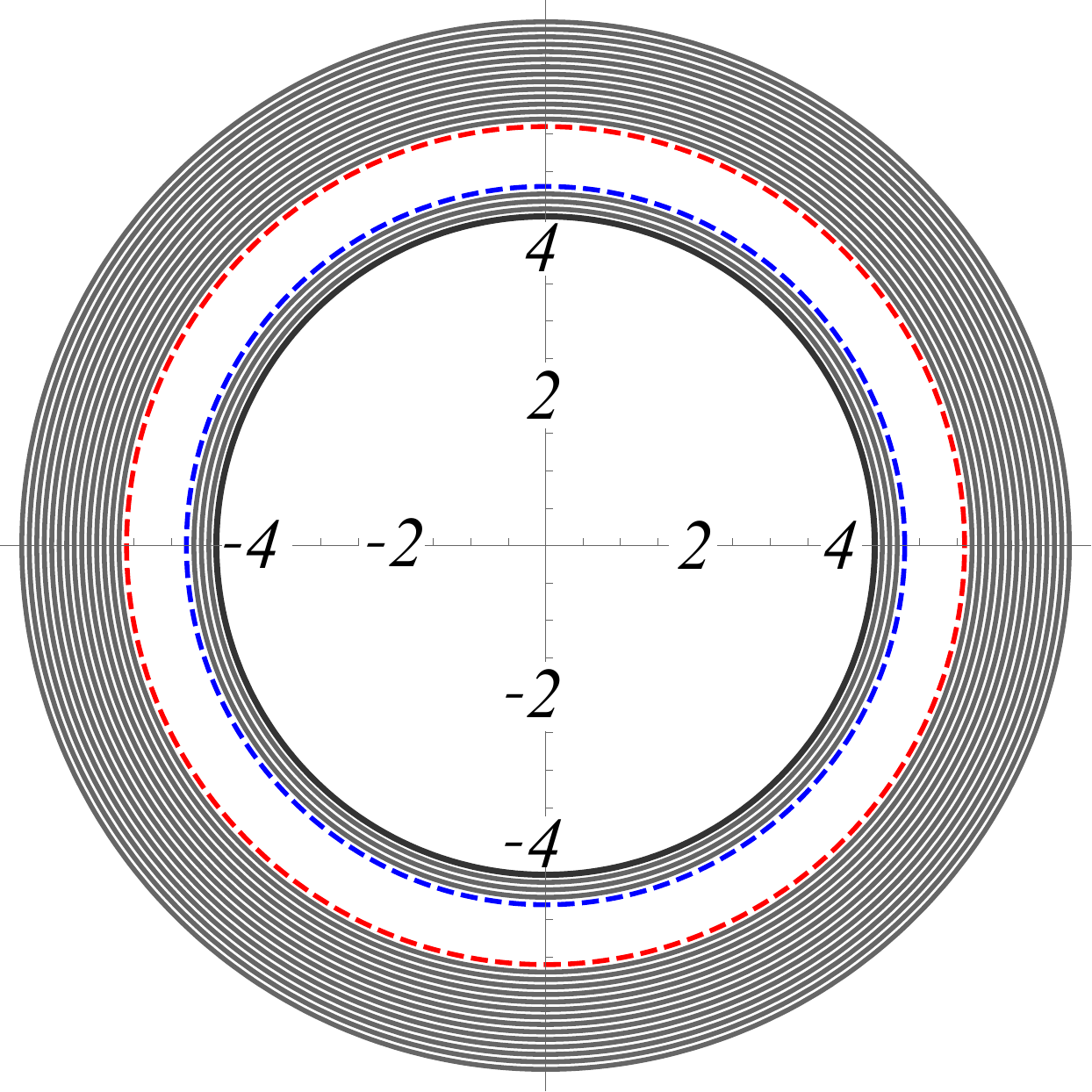}
             \includegraphics[width=0.32\textwidth]{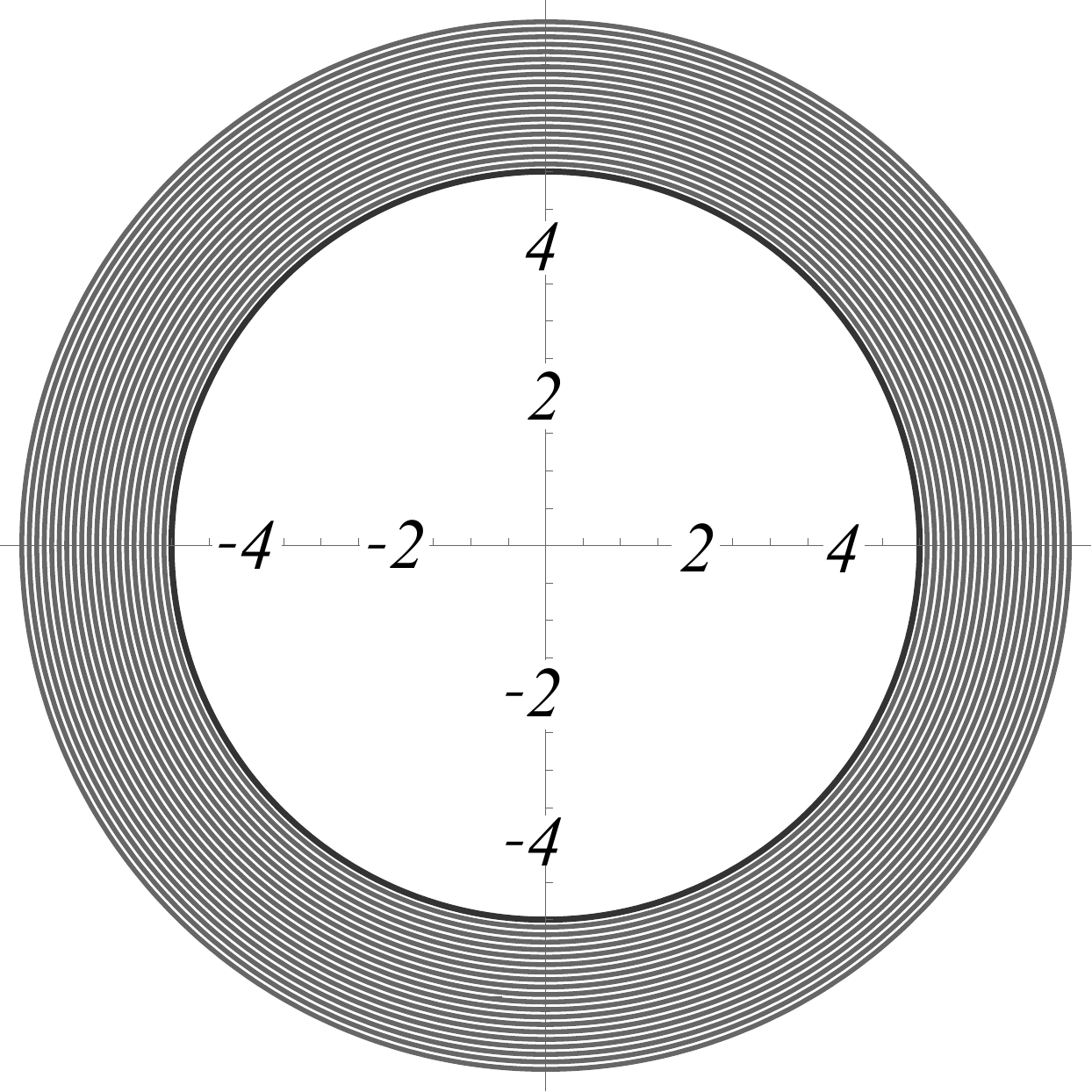} \\[2mm]
              $a) \,\,\,\gamma=0.9$ \hspace{3cm} $b) \,\,\,\gamma=0.455$ \hspace{3cm} $c) \,\,\,\gamma=0.4$ \\[1mm]
           		\end{tabular}}
 \caption{\label{fig:disk}\small The possible qualitatively different structures of the accretion disk that can occur for a) weakly naked, and b), c) strongly naked singularities. The curvature singularity $r_{cs}$ is represented by the black line, while the inner and outer marginally stable orbits $r_{-}$ and $r_{+}$ are depicted by the blue and the red dashed lines, respectively. We give also the location of the photon sphere by the orange dashed line. }
\end{figure}

We can expect that the different physical regimes would have also impact on the behavior of the timelike geodesics in the Janis-Newman-Winicour spacetime. The structure of the timelike circular geodesics in the equatorial plane was investigated in detail in \cite{Chowdhury:2012}, however in a different parametrization of the solution. Performing the standard analysis of the geodesic motion in spherically symmetric spacetimes by associating an effective potential to it and evaluating its extrema, it can be shown that the stability of the circular geodesics in the equatorial plane is governed by the equation

 \begin{equation}\label{ISCO_eq}
r^2\gamma^2 - 2r\gamma(3\gamma + 1) + 2(2\gamma^2 + 3\gamma + 1) =0.
\end{equation}
It possesses the roots

\begin{equation}
r_\pm = \frac{1}{\gamma}\left(3\gamma+1 \pm \sqrt{5\gamma^2 -1}\right),
\end{equation}
which delimit the regions of stability as the stable circular orbits' radial positions should satisfy  $r>r_{+}$, or $r_{cs}<r<r_{-}$, when the last inequality is reasonable. In the weakly naked case only $r_{+}$ lies outside the curvature singularity $r_{cs}$. Therefore, the region of stable circular orbits is characterized by the same structure as for the Schwarzschild black hole. They begin at $r_{ISCO} = r_{+}$ corresponding to the innermost stable circular orbit and span to infinity. For strongly naked singularities real roots of eq. ($\ref{ISCO_eq}$) exist only for values of the solution parameter $1/\sqrt{5} < \gamma < 0.5$. In this range it is satisfied that $r_{cs}<r_{-}$, so the stable circular orbits are distributed in two disconnected annular regions limited by the  marginally stable orbits $r_\pm$. For $\gamma < 1/\sqrt{5}$ all the circular geodesics in the equatorial plane are stable. We illustrate the three different structures of the accretion disk in fig. $\ref{fig:disk}$.

\begin{figure*}[h!]
    \centering
    \begin{subfigure}[t]{0.7\textwidth}
        \includegraphics[width=\textwidth]{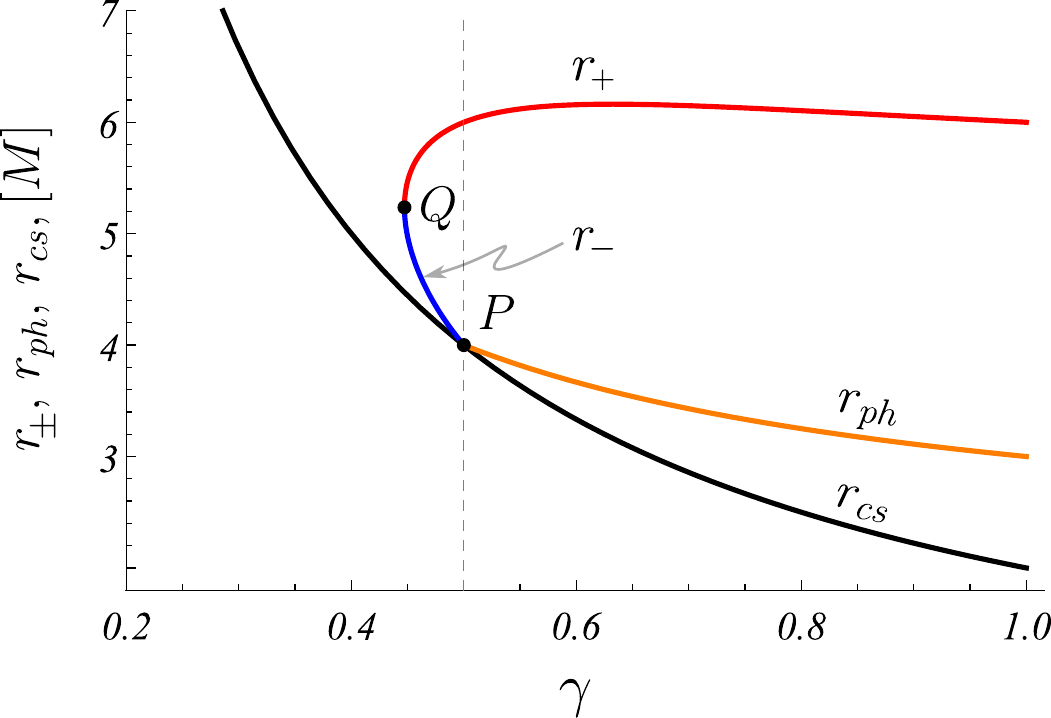}
           \end{subfigure}
           \caption{\label{fig:ISCO}\small Location of the curvature singularity $r_{cs}$ (black curve), the photon sphere $r_{ph}$ (orange curve), and the inner and the outer marginally stable orbits $r_{-}$ and $r_{+}$ (in blue and red, respectively) as a function of the solution parameter $\gamma$. For $\gamma = 1/2$ (point $P$)  the positions of the curvature singularity, the photon sphere, and the inner marginally stable orbit coincide. For lower values of $\gamma$  no photon sphere exists, and a second region of stability of the circular geodesics develops. For $\gamma = 1/\sqrt{5}$ (point $Q$) the two marginally stable orbits merge, and for $\gamma < 1/\sqrt{5}$ all the circular geodesics are stable.}
\end{figure*}

The transition between the qualitatively different configurations of the accretion disk can be understood by examining the behavior of the location of the curvature singularity  $r_{cs}$, and the marginally stable circular orbits $r_{\pm}$ when varying $\gamma$ (see fig. $\ref{fig:ISCO}$). We can also consider the radial position of the photon sphere, which is given by

\begin{equation}
r_{ph} = (2\gamma + 1)M/\gamma.
\end{equation}

We start with the Schwarzschild solution with $\gamma =1$ and decrease the value of $\gamma$ in the limits $1/2 < \gamma < 1$. In this range both the curvature singularity and the photon sphere move to larger values of the radial coordinate. However, the position of the curvature singularity grows faster, and for $\gamma = 1/2$ they merge at $r=4M$. For lower values of $\gamma$ $r_{ph}<r_{cs}$ is satisfied, consequently no photon sphere exists. For $\gamma < 1/2$ the position of the inner marginally stable circular orbit $r_{-}$ becomes larger than $r_{cs}$. It continues to increase as, at the same time, the position of the outer marginally stable circular orbit $r_{+}$ decreases, causing the separation between the two disconnected parts of the accretion disk to shrink. For $\gamma = 1/\sqrt{5}$ the two orbits coincide, and for smaller values of $\gamma$ no marginally stable orbits exist. Then, the stable circular orbits are possible for any radial distance, filling up the whole spacetime.

\section{Optical appearance of a thin accretion disk around the strongly naked JNW singularity}

We consider an accretion disk around the strongly naked Janis-Newman-Winicour singularity in the thin disk approximation when the disk height is assumed to be negligible. Then, the accretion process is modeled by a collection of particles, which move on stable circular orbits in the equatorial plane and emit radiation isotropically. In the case of the strongly naked JNW singularity we can distinguish two different types of possible distributions of the stable circular orbits, which were described in the previous section. Depending on the scalar charge to mass ratio, the accretion disk can extend continuously through the whole spacetime up to the singularity, or consist of two disconnected regions delimited by a couple of marginally stable orbits (see fig. $\ref{fig:disk}$). In this work we study the second case, which corresponds to values of the solution parameter in the range $1/\sqrt{5}<\gamma<1/2$. Then, the accretion disk is still parameterized by some characteristic distances, representing its edges, similar to the ISCO in the black hole case.

In order to obtain the observable image of the accretion disk we use the approach described in our previous work \cite{Nedkova:2019}. The optical appearance of each circular orbit is calculated , as seen by an observer located at the asymptotic infinity. For the purpose we apply two independent procedures by taking advantage of the spherical symmetry of the spacetime, or assuming only stationarity and axial symmetry. We visualize a certain portion of the disk up to $r = 30M$, which contains the region, where the gravitational interaction is strong, and most of the relativistic processes occur, and adopt as an effective measure of the asymptotic infinity the observer position $r_{obs}= 5000M$.

The spherical symmetry allows to develop a semi-analytical scheme for computing the observable image of the spherical orbits, which involves only numerical integration. We consider the constraint equation for the null geodesics $g_{\mu\nu}{\dot x}^{\mu}{\dot x}^{\nu} =0$, where the dot denotes differentiation with respect to an affine parameter. For the Janis-Newman-Winicour spacetime it takes the form

\begin{equation}
{\dot r}^2 + \frac{L^2}{r^2}\left(1-\frac{2M}{\gamma r}\right)^{2\gamma -1}=E^2,
\end{equation}
where we introduce the integrals of motion $E$ and $L$, representing the photon's specific energy  and angular momentum, respectively. We can further consider
the geodesic equation for the azimuthal angle

\begin{equation}
{\dot \phi}= \frac{L}{r^2}\left(1-\frac{2M}{\gamma r}\right)^{\gamma -1},
\end{equation}
and  obtain its variation along the photon trajectory as a function of the radial coordinate

\begin{equation}\label{delta_phi_int}
\Delta\phi = \int{\frac{dr}{r^2\left(1-\frac{2M}{\gamma r}\right)^{1-\gamma}\sqrt{\frac{1}{b^2} -\frac{1}{r^2}\left(1-\frac{2M}{\gamma r}\right)^{2\gamma -1}}}},
\end{equation}
by means of the impact parameter $b=L/E$, which depends on the ratio of the specific angular momentum and the energy on the geodesic. In order to construct the observable image of a circular orbit we compute numerically the integral between the orbit's position, which we denote by $r_{source}$, and the location of an observer at spatial infinity, which we assume to correspond to $r_{obs} = 5000M$. Without any limitations we can choose the value of the azimuthal angle at the observer's position to be equal to $\phi_{obs} =0$. In this way its value at the photon's emission point corresponds to the total azimuthal variation along the geodesic trajectory.

The next step is to obtain the projection of the photon trajectory on the observers's sky. For the purpose we should introduce a local reference frame at the observer's location. If we consider a static spherically symmetric metric in the form
\begin{equation}\label{metric_st}
ds^2=g_{tt}\,dt^2+g_{\phi\phi}\,d\phi^2+g_{rr}\,dr^2+g_{\theta\theta}\,d\theta^2\,,
\end{equation}
the natural othonormal tetrad adapted to the spacetime symmetries is given by

\begin{eqnarray}\label{ref_frame}
\hat{e}_{(\theta)}&=&\frac{1}{\sqrt{g_{\theta\theta}}}\partial_\theta,\quad~~~ \hat{e}_{(r)}=\frac{1}{\sqrt{g_{rr}}}\partial_r,\\
\hat{e}_{(\phi)}&=&\frac{1}{\sqrt{g_{\phi\phi}}}\partial_\phi, \qquad~~~\hat{e}_{(t)}=\frac{1}{\sqrt{-g_{tt}}}\,\partial_t, \nonumber
\end{eqnarray}
where the radial vector $\hat{e}_{(r)}$ corresponds to the direction of the compact object. In this local reference frame we can introduce two celestial angles $\xi$ and $\eta$, which serve as the coordinates of the observable image.  The celestial coordinate $\xi$ is defined as the angle between the photon trajectory and the compact object direction $\hat{e}_{(r)}$, while $\eta$ is the angle between the plane, in which the geodesic lies,  and the  basis vector $\hat{e}_{(\phi)}$. The observer is located at the inclination angle $i$ with respect to the normal direction to the orbital plane of the emitting particle. The projection geometry is described in detail in \cite{Muller:2009} (see fig. 7 there).

The spherical symmetry allows to derive a relation between the value of the azimuthal angle $\phi$ at the photon emission point,  the celestial angle $\eta$, and the inclination angle $i$ \cite{Luminet:1979},  \cite{Muller:2009}

\begin{eqnarray}\label{delta_phi}
\cos\phi = - \frac{\sin\eta\tan i}{\sqrt{\sin^2\eta\tan^2 i + 1}}, \quad~~~ \sin\phi = \frac{1}{\sqrt{\sin^2\eta\tan^2 i + 1}},
\end{eqnarray}
where $\phi$ is assumed to take values in the range $[0, \pi)$. On the other hand, using the orthonormal tetrad ($\ref{ref_frame}$) we can express the celestial angle $\xi$ by means of the impact parameter $b$ on the geodesic as

\begin{equation}\label{xi}
\xi =\arcsin{\frac{b}{r_{obs}}\left(1 - \frac{2M}{\gamma r_{obs}}\right)^{\gamma - \frac{1}{2}}}.
\end{equation}

The two representations of the azimuthal angle $\phi$ given by eqs. ($\ref{delta_phi_int}$) and ($\ref{delta_phi}$) provide a condition for the impact parameter on the geodesic, which should be satisfied for photon emission points, which can be observed at inclination angle $i$ and celestial coordinate $\eta$
\begin{equation}\label{eta_b}
\int^{r_{obs}}_{r_{source}}{\frac{dr}{r^2\left(1-\frac{2M}{\gamma r}\right)^{1-\gamma}\sqrt{\frac{1}{b^2} -\frac{1}{r^2}\left(1-\frac{2M}{\gamma r}\right)^{2\gamma -1}}}} = -\arccos{ \frac{\sin\eta\tan i}{\sqrt{\sin^2\eta\tan^2 i + 1}}}.
\end{equation}
By taking advantage of eq. ($\ref{xi}$), it can be further transformed into a relation between the two celestial coordinates $\xi= \xi(\eta)$. Practically, for a particular inclinational angle $i$ and orbit position $r_{source}$ we scan all the observational angles $\eta \in [0,2\pi]$ and obtain the corresponding impact parameters $b$, which satisfy ($\ref{eta_b}$). Using eq. ($\ref{xi}$) this procedure defines a curve $\xi= \xi(\eta)$ on the observer's sky, which represents  the image of the circular orbit located at the radial coordinate $r_{source}$  observable at the inclination angle $i$.

The described scheme does not provide all the possible images of a particular circular orbit, since the azimuthal angle is restricted to the range $\phi \in [0, \pi)$. In general, there exist trajectories, which are deflected at larger angles, or perform several circles around the compact object before reaching the observer. Therefore, eq. ($\ref{eta_b}$) should be generalized to include all the possible cases

\begin{equation}\label{orbit_im}
\int^{r_{obs}}_{r_{source}}{\frac{dr}{r^2\left(1-\frac{2M}{\gamma r}\right)^{1-\gamma}\sqrt{\frac{1}{b^2} -\frac{1}{r^2}\left(1-\frac{2M}{\gamma r}\right)^{2\gamma -1}}}} = k\pi -\arccos{ \frac{\sin\eta\tan i}{\sqrt{\sin^2\eta\tan^2 i + 1}}},
\end{equation}

\noindent
by introducing an image order given by the non-negative integer $k$, which classifies the qualitatively different cases that can arise. The simplest case $k=0$ gives rise to the so called direct images, which are generated by photons emitted in the direction above the equatorial plane. Trajectories  with $k=1$ form secondary images, corresponding to photons heading in the direction below the equatorial plane. Null geodesics of higher order $k\geq2$ include several circles around the compact object before reaching the observer. When we increase the order $k$ in spacetimes possessing a photon sphere such geodesics produce images which asymptotically approach the image of the photon sphere.

By using the above procedure we can obtain the full set of images of any circular orbit. In addition we compute the disk image in a second way by applying a fully numerical ray-tracing scheme. It is valid for any stationary and axisymmetric spacetime since it assumes only two integrals of motion, connected with the photon's energy and angular momentum, and does not make use of the spherical symmetry of the solution. In case of stationary and axisymmetric spacetime it is convenient to parameterize the projection of the geodesics on the observer's sky by a different set of celestial coordinates. Following \cite{Cunha:2016a} we introduce two celestial angles  $\alpha \in [0,\pi]$ and $\beta \in [-\pi/2,\pi/2]$, which are related to the photon's 4-momentum as

\begin{eqnarray}\label{p_alpha}
&&p_\theta=\sqrt{g_{\theta\theta}}\sin\alpha,\qquad\qquad\,\,\, p_\phi =L =\sqrt{g_{\phi\phi}}\sin\beta\,\cos\alpha, \nonumber \\[2mm]
&&p_r=\sqrt{g_{rr}}\cos\beta\,\cos\alpha,\qquad p_t = -E.
\end{eqnarray}

\noindent
When the observer is located at the asymptotic infinity, the coordinates $\alpha$ and $\beta$ are related to the observational angles $\eta$ and $\xi$, which we used in the previous procedure as $\alpha = \xi\cos\eta$, $\beta = \xi\sin\eta$. For fixed inclination angle of the observer $\alpha$ and $\beta$ can be interpreted as initial data on the geodesic and we can integrate numerically the photon's trajectory backwards to its emission point. Then, the construction of the accretion disk image transforms into the following  problem. For any $\alpha$ and $\beta$ in their relevant ranges we try to obtain photon trajectories which pass through some stable circular orbit in the equatorial plane, i.e. through a point with coordinates $\theta = \pi/2$ and $r \in (r_{cs}, r_{-}]$ or $r \in [r_{+}, 30M]$, where $r_{\pm}$ are the two marginally stable orbits delimiting the disk edges, and we have chosen to visualize the accretion disk up to $r=30M$. The set of celestial angles $\alpha$ and $\beta$, for which a solution of the above problem exists, forms the image of the accretion disk on the observer's sky. The numerical scheme provides the image of the whole accretion disk simultaneously without giving information about the optical appearance of the separate circular orbits. In this respect the semi-analytical procedure is valuable since it gives intuition how the image of each particular circular orbits is produced, and how high image orders can be observed.

\begin{figure}[h!]
\setlength{\tabcolsep}{ 0 pt }{\footnotesize\tt
		\begin{tabular}{ c }
           \includegraphics[width=\textwidth]{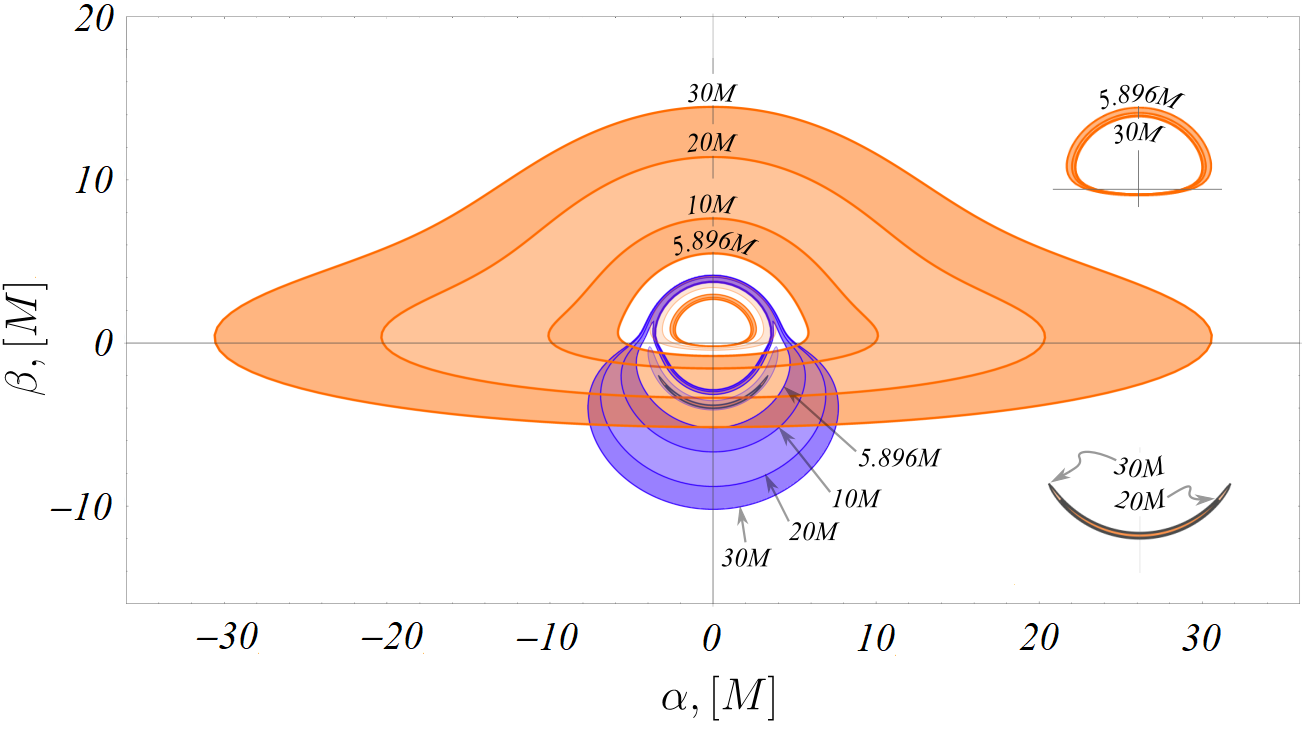}\\[1mm]
        \end{tabular}}
 \caption{\label{fig:IsoR1}\small Optical appearance of the strongly naked Janis-Newman-Winicour singularity with a thin accretion disk for $\gamma = 0.48$. The observer is located at $r=5000M$ and the inclination angle $i=80^\circ$. The direct image of the accretion disk is depicted in orange, the secondary image with $k=1$ is in blue, while the higher order images are presented in black contour.  The two marginally stable orbits, which delimit the inner and the outer disk, are located at $r_{-}= 4.271M$ and $r_{+} = 5.896M$. In the side images we present a zoom-out of the second direct image of the outer disk (upper right), which forms a ring in the central region, and its image of order $k=2$ (down right).  The image of the inner disk is given in very light orange and very light blue in the main picture. }
\end{figure}

We apply the described computational methods to obtain the optical appearance of the strongly naked Janis-Newman-Winicour singularity (see fig. $\ref{fig:IsoR1}$). We choose the solution parameter to take the value $\gamma=0.48$ noting that the results are representative for any value of the solution parameter in the range $1/\sqrt{5}<\gamma<1/2$, since the observable images share the same qualitative features. The observer is assumed to be located at $r_{obs}=5000M$, and at the inclination angle $i= 80^\circ$. The images for large inclination angles are most informative since the observable deformations of the accretion disk become more pronounced with the increase of the inclination angle, and the relativistic effects get more clearly distinguished.

In fig. $\ref{fig:IsoR1}$  two images are superposed generated by the numerical and the semi-analytical procedures. The numerical procedure visualizes the optical appearance of a continuous distribution of circular orbits starting from a small neighbourhood of the curvature singularity $r_{cs}$ and extending to the inner marginally stable orbit $r_{-}$ combined with the set of orbits lying between the outer  marginally stable orbit $r_{+}$ and  $r=30M$. For simplicity we will refer  to the two regions as inner and outer accretion disks. For the solution parameter $\gamma=0.48$ the characteristic radial coordinates delimiting the inner and the outer disks take the values $r_{-}= 4.271M$ and $r_{+} = 5.896M$, and the curvature singularity is located at $r_{cs}=4.166667M$. In all the computations we avoid the curvature singularity by excluding a neighbourhood with a radius $\delta r = r_{cs} + 10^{-10}M$ around it.

\begin{figure}[t!]
    		\setlength{\tabcolsep}{ 0 pt }{\small\tt
		\begin{tabular}{ cc}
           \includegraphics[width=0.6385\textwidth]{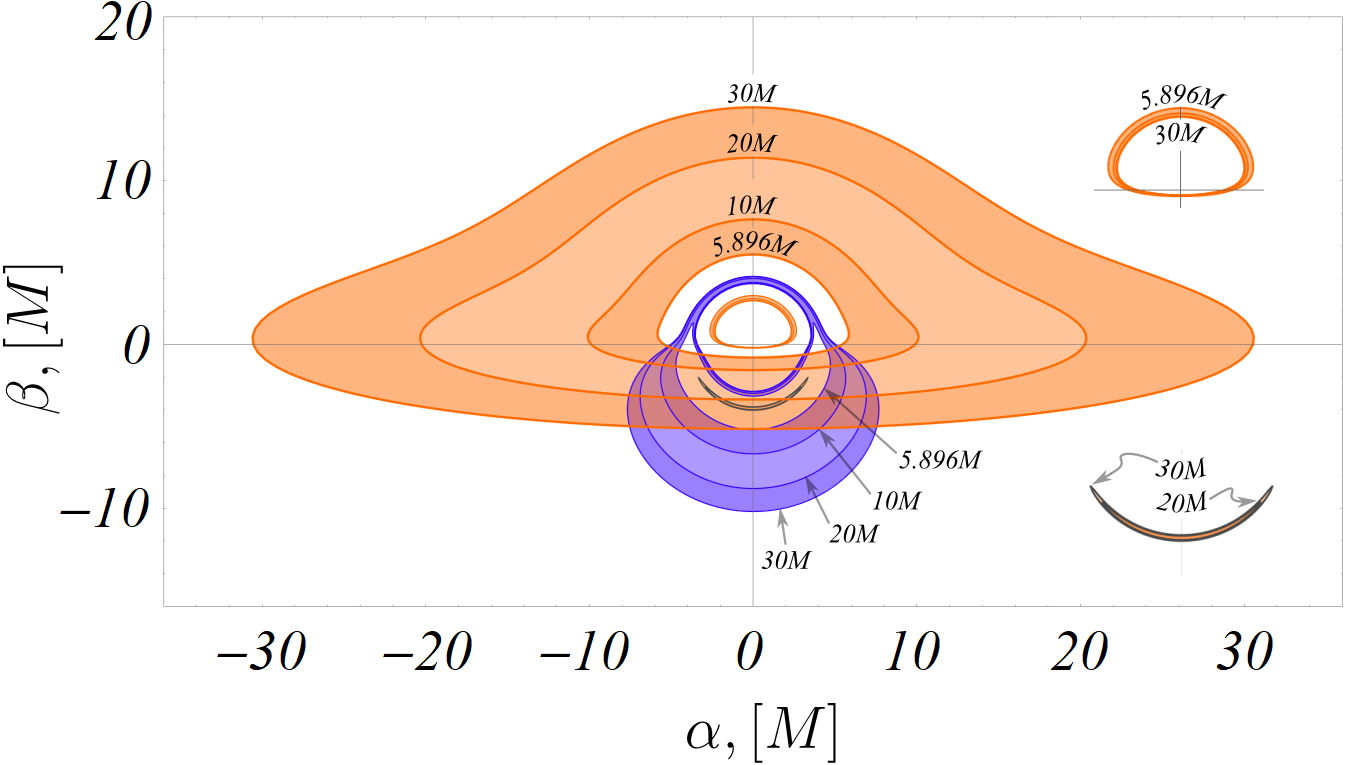}
           \includegraphics[width=0.3547\textwidth]{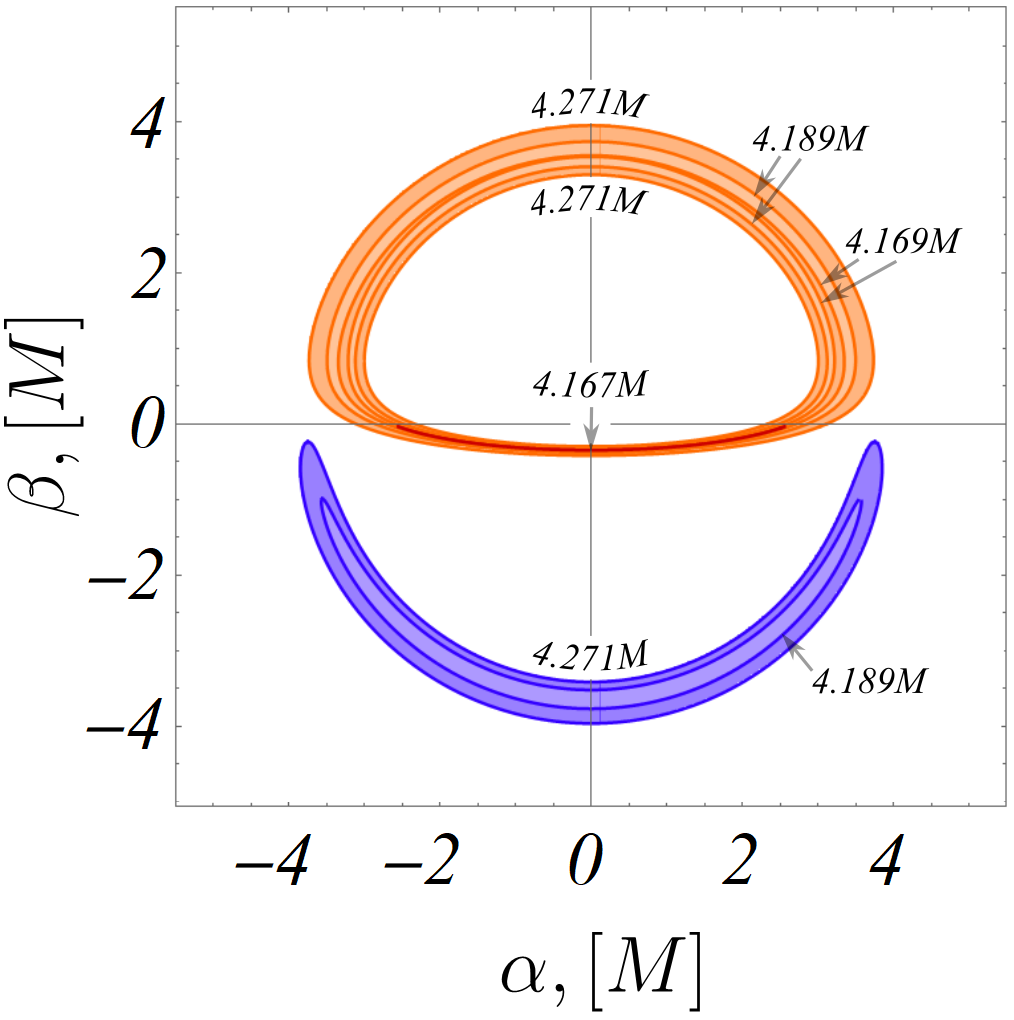} \\[1mm]
           \hspace{3.0cm}  $a)$ \hspace{7.0cm}  $b)$
        \end{tabular}}
 \caption{\label{fig:IsoR2}\small Optical appearance of the outer (left) and the inner (right) accretion disks around the strongly naked Janis-Newman-Winicour singularity with $\gamma=0.48$. The direct image is in orange, the secondary image with $k=1$ is in blue, while the higher order images are presented in black contour.  The inner marginally stable orbit is located at $r_{-}= 4.271M$ , while the outer one corresponds to $r_{+} = 5.896M$. The inner disk is visualized up to a small neighbourhood $\delta r = 10^{-10}M$ of the curvature singularity, which is located at $r_{cs}=4.166667M$.}
\end{figure}

\begin{figure}[h!]
\setlength{\tabcolsep}{ 0 pt }{\footnotesize\tt
		\begin{tabular}{ c }
           \includegraphics[width=\textwidth]{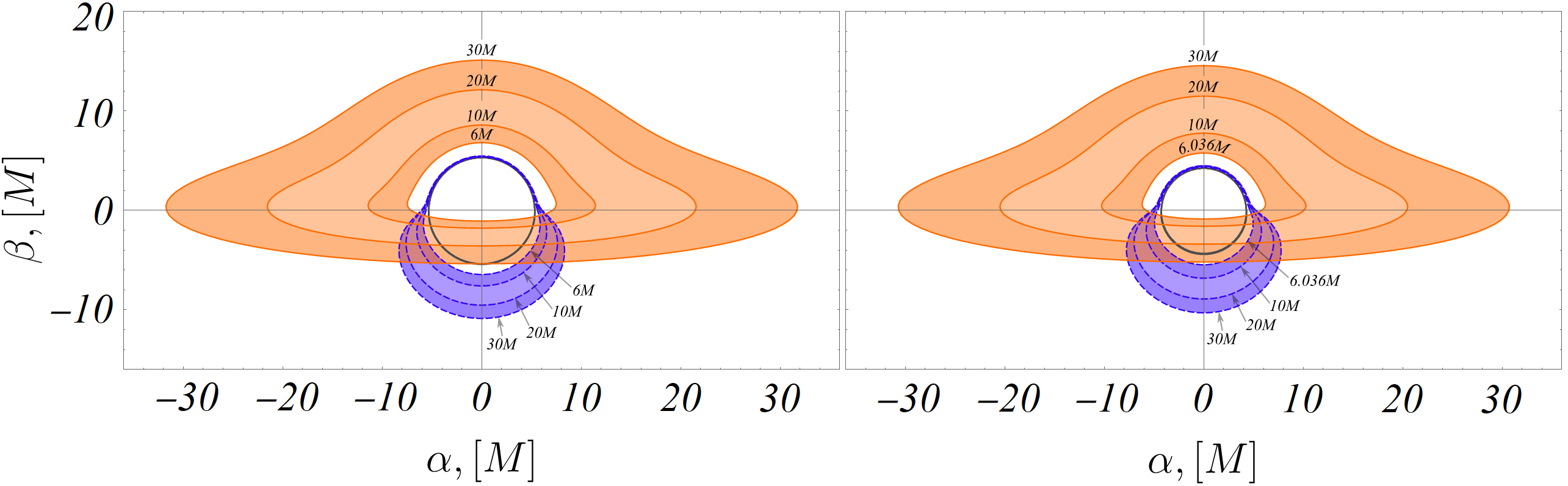}\\[1mm]
        \end{tabular}}
 \caption{\label{fig:IsoR3}\small Optical appearance of a thin accretion disk around the Schwarzschild black hole (left) and the weakly naked Janis-Newman-Winicour singularity (right) with solution parameter $\gamma = 0.51$. The observer is located at $r=5000M$ and inclination angle $i=80^\circ$. The black circle represents the optical appearance of the photon sphere. We use the same conventions as in fig. $\ref{fig:IsoR1}$. }
\end{figure}

Using the semi-analytical procedure we obtain the images of a discrete set of circular orbits, which are represented by solid lines in the plots with indications about the orbit's radial position attached to them. In order to be able to differentiate between the images of different order, we illustrate the direct images ($k=0$) in orange, the secondary images with $k=1$ in blue, and the higher order images in black contour. For the Janis-Newman-Winicour naked singularity with $\gamma =0.48$ the highest order of the image of any orbit, which can be observed at inclination angle $i=80^\circ$ is $k=2$. We give an explanation of this property in the following analysis in section 4. The optical appearance of the accretion disk presented in fig. $\ref{fig:IsoR1}$ contains the images of the both disconnected sets of stable circular orbits representing the inner and outer disks. In order to make the interpretation of the image more intuitive we illustrate the separate contributions of the inner and outer disks in fig. $\ref{fig:IsoR2}$. For comparison we also provide the optical appearance of a thin accretion disk and a weakly naked Janis-Newman-Winicour singularity in fig. $\ref{fig:IsoR3}$.

The most distinctive feature in the optical appearance of the thin accretion disk around a strongly naked singularity is the formation of several types of rings in the central part of the image. In the case of the Schwarzschild black hole and the weakly naked singularity this region in the image is dark and corresponds to the compact object's shadow. Thus, the observation of a set of central rings may be used to distinguish naked singularities from black holes in the imaging experiments. For the strongly naked singularity the ring structure consists of several types of images. The darker orange ring represents a direct image of the outer accretion disk. The outer disk, which expands between the outer marginally stable orbit $r_{+} = 5.896M$ and the truncation radius $r=30M$, gives rise to two disconnected direct images. One of them corresponds to the usual disk image familiar from the Schwarzschild black hole, while the other one appears as a ring in the region enclosed by the outer marginally stable orbit. In addition, in the central region we observe a second ring with larger radius, which we depict in very light orange. It corresponds to the image of the inner accretion disk, which spans from the curvature singularity to the inner marginally stable orbit $r_{-} = 4.271M$. The third central ring depicted in blue, which partially overlaps with the previous one, is a secondary image with $k=1$ of the outer accretion disk. We see that in the case of the Schwarzschild black hole and the weakly naked singularity we have only the secondary image. However, it is compressed to a very small neighbourhood of the image of the photon sphere, and is practically unobservable.

\section{Formation of the accretion disk image}

Using the semi-analytical procedure for constructing the accretion disk image we can explain the mechanism of formation of the central ring images, as well as the appearance of a double  direct image of the outer accretion disk. For the purpose we should examine in more detail the structure of eq. ($\ref{orbit_im}$). The integral on the left-hand side, which defines the variation of the azimuthal angle, can take different form depending on the photon trajectory. There exist two types of trajectories - such possessing no radial turning point, and others, which pass through a single radial turning point $r_0$. If a turning point is present, the integral is represented as a sum of two integrals in order to describe the photon motion

\begin{eqnarray}\label{orbit_im1}
\Delta\phi &=& \int^{r_{source}}_{r_0}{\frac{dr}{r^2\left(1-\frac{2M}{\gamma r}\right)^{1-\gamma}\sqrt{\frac{1}{b^2} -\frac{1}{r^2}\left(1-\frac{2M}{\gamma r}\right)^{2\gamma -1}}}} \nonumber \\[2mm]
&+& \int^{r_{obs}}_{r_0}{\frac{dr}{r^2\left(1-\frac{2M}{\gamma r}\right)^{1-\gamma}\sqrt{\frac{1}{b^2} -\frac{1}{r^2}\left(1-\frac{2M}{\gamma r}\right)^{2\gamma -1}}}}.
\end{eqnarray}

\noindent
Using the same property we can obtain a relation between the impact parameter of the geodesic and the location of the turning point $r_0$

\begin{eqnarray}\label{D_r}
b =  r_0\left(1-\frac{2M}{\gamma \, r_0}\right)^{\frac{1}{2}-\gamma}.
\end{eqnarray}
Thus, geodesics possessing a turning point can be parameterized by its position instead of the impact parameter. The relation is a monotonically increasing function, and the value of the turning point position should be always lower than the radial position of the photon emission point $r_{source}$. Therefore, there exists a maximum value $b_{max}$, which the impact parameter can take, which is obtained from ($\ref{D_r}$) in the limit when $r_0$ approaches $r_{source}$

\begin{eqnarray}
b_{max} =  r_{source}\left(1-\frac{2M}{\gamma \, r_{source}}\right)^{\frac{1}{2}-\gamma}.
\end{eqnarray}
We can consider the variation of the azimuthal angle corresponding to the trajectory with the maximum value of the impact parameter, which we denote by $\phi_{crit}$. This critical value defines the transition point between the two possible types of photon trajectories. For given positions of the source and the observer, trajectories with azimuthal variation $\phi \leq \phi_{crit}$ possess no radial turning point, while those with $\phi > \phi_{crit}$ pass through a turning point.

\begin{figure}[t!]
    		\setlength{\tabcolsep}{ 0 pt }{\small\tt
		\begin{tabular}{ cc}
           \hspace{0.5cm}\includegraphics[width=0.4\textwidth]{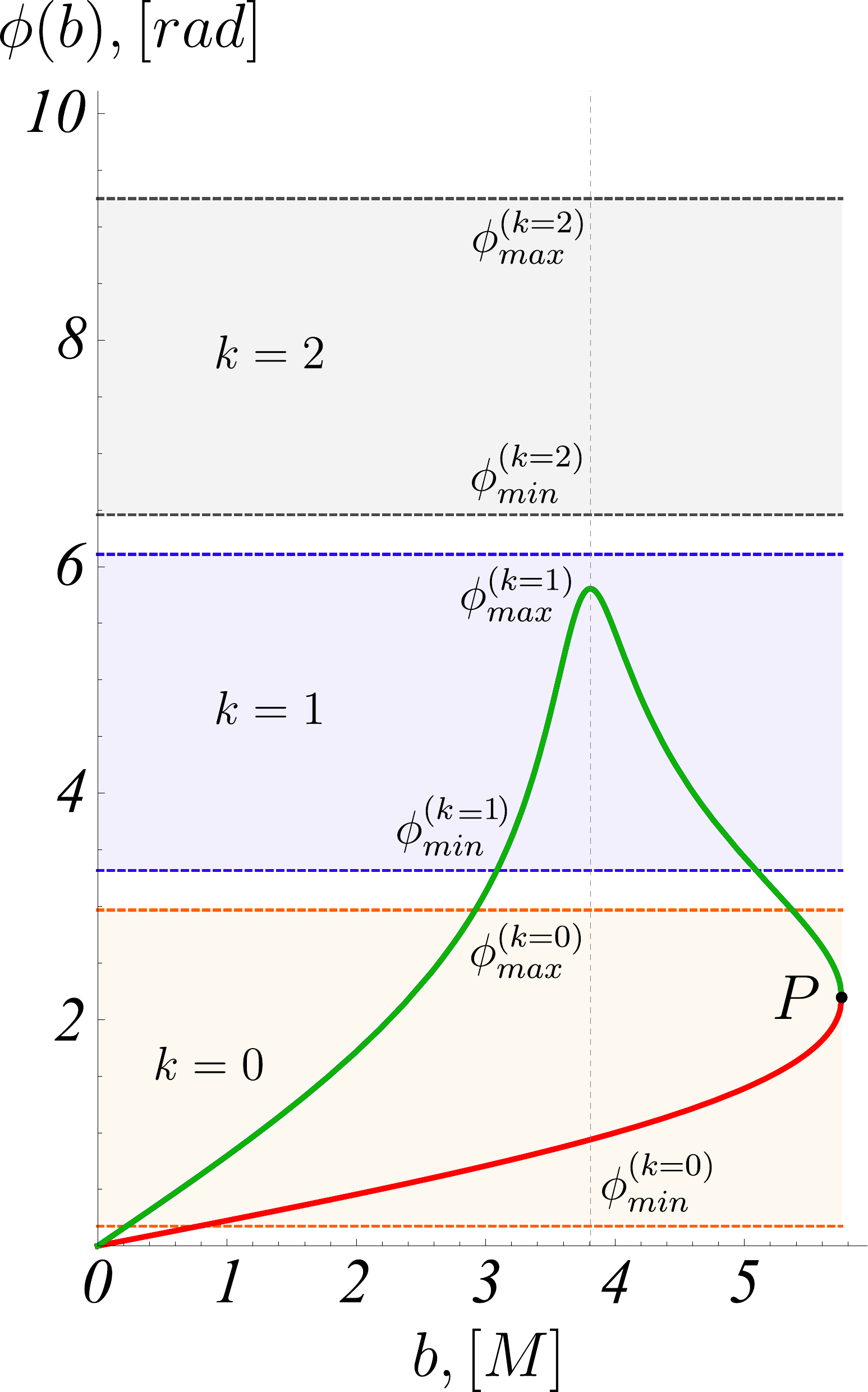}  \hspace{1cm}
		   \includegraphics[width=0.4\textwidth]{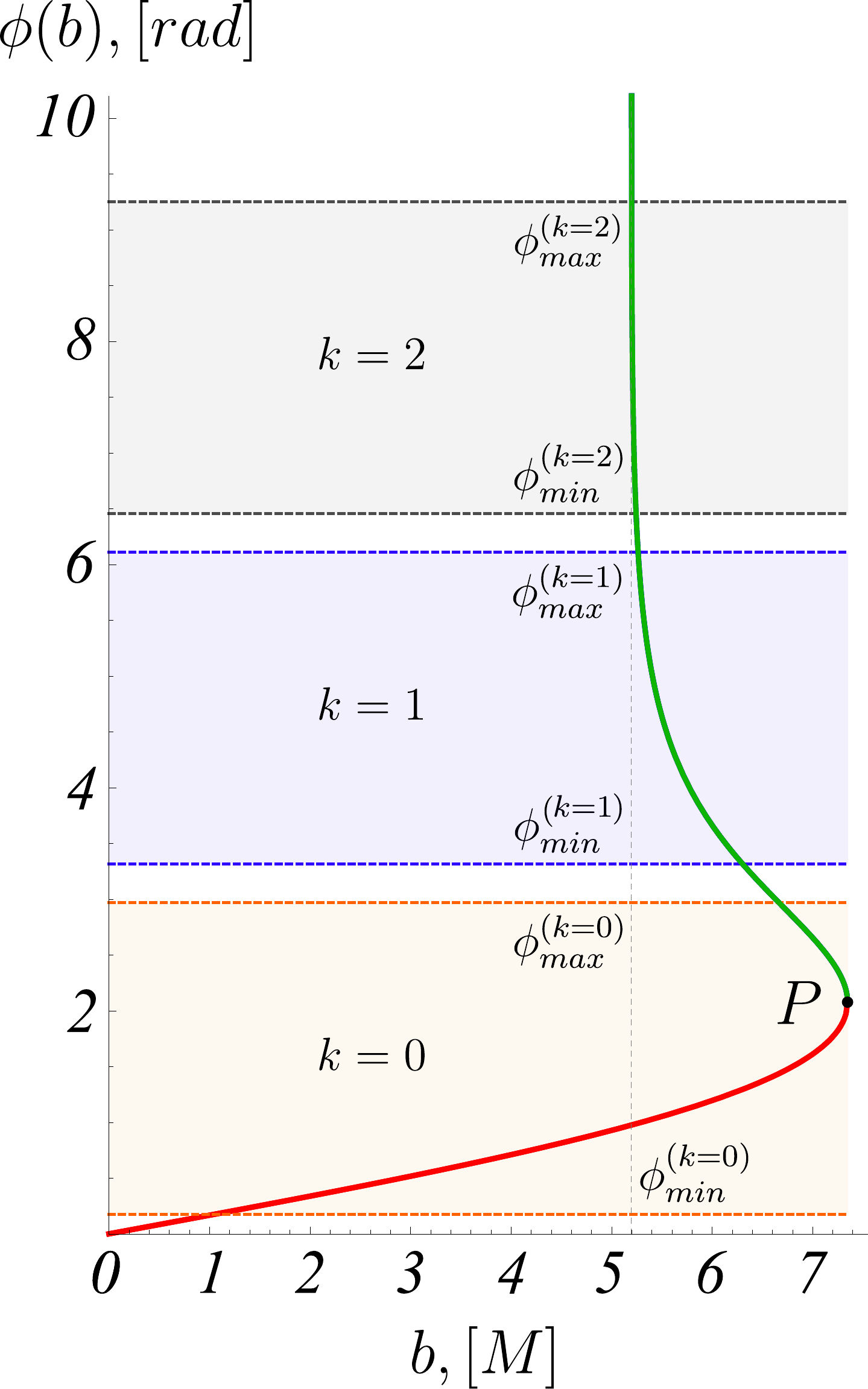} \\[1mm]
           \hspace{1.2cm}  $a)$ \hspace{7cm}  $b)$
        \end{tabular}}
 \caption{\label{fig:Int_r1}\small Image formation diagrams for the outer marginally stable orbit $r_{+} = 5.896M$ of the strongly naked singularity with $\gamma=0.48$ a) and for the ISCO of the Schwarzschild solution b). The observer is located at $r=5000M$ and at the inclination angle $i=80^\circ$. The point P represents the maximum value for the impact parameter $b_{max}$, for which we reach the critical azimuthal angle $\phi_{crit}$. The main distinction between the two cases is that for the strongly naked singularity the green curve is limited, possessing a single local maximum, while for the Schwarzschild solution it diverges at the impact parameter corresponding to the photon sphere. This leads to the possibility that two disconnected portions of the green and the red curves intersect a certain observational window, which results in the formation of two disconnected images a given order. }
\end{figure}

Following these  arguments we can describe graphically the solutions of eq. ($\ref{orbit_im}$) as presented in fig. $\ref{fig:Int_r1}$. For the strongly naked singularity with $\gamma=0.48$ we consider all the photon trajectories emitted from the outer marginally stable orbit at $r_{source}= r_{+}=5.896M$ and reaching an observer at $r_{obs} = 5000M$, which can be observed at inclination angle $i=80^\circ$. Then, for every image order $k$ the right-hand side of eq. ($\ref{orbit_im}$) defines a minimal and a maximal observable value of the azimuthal angle $\phi^{k}_{min}$ and $\phi^{k}_{max}$, which correspond through eq. ($\ref{delta_phi}$) to the minimal and maximal values of the celestial coordinate $\eta$, respectively. Thus, for every $k$ we can define an observational window $\Delta\phi^{k} = \phi^{k}_{max}- \phi^{k}_{min}$, which determines the possible azimuthal variation for the observable trajectories. These intervals are depicted in orange, blue and grey in fig. $\ref{fig:Int_r1}$ according to the image order. We further represent the integral on the left-hand side as a function of the impact parameter by a green and a red curve, respectively, for the two cases when there exists or doesn't exist a turning point.  The two curves intersect at the maximal value of the impact parameter $b=b_{max}$ for azimuthal angle $\phi = \phi_{crit}$. Thus, for any value of the impact parameter $b\in(0, b_{max})$ we have a solution of the integral $\phi^{(1)}< \phi_{crit}$ without turning point belonging to the red curve, and another one $\phi^{(2)}> \phi_{crit}$, which corresponds to a trajectory with a turning point and belongs to the green curve.

The full image of a given circular orbit is determined by all the possible intersections of the green and the red curves with any of the observational windows $\Delta\phi^{k}$. Depending on the structure of the intersections we can outline the following rules for describing the images that will arise:

\begin{itemize}
\item {} An intersection with an observational window $\Delta\phi^{k}$ results in the formation of an image of order $k$.
\item{} An intersection of two disconnected portions of the green and the red curves (except for the point $b=0$) with an observational window $\Delta\phi^{k}$ leads to the formation of two disconnected images of order $k$, which we call a double image of order $k$.
\item{}  An intersection of the red or the green curve with an observational window $\Delta\phi^{k}$, which does not span through the whole range of $\Delta\phi^{k}$, results in an image, which does not close around the origin of the celestial coordinate system. This feature arises since the lack of solutions for some azimuthal angles $\phi \in \Delta\phi^{k}$ means that for some of the celestial angles $\eta\in[0,2\pi]$ there exists no image of the given type.
\end{itemize}

\begin{figure}[h!]
    		\setlength{\tabcolsep}{ 0 pt }{\small\tt
		\begin{tabular}{ cc}
           \hspace{0.5cm}\includegraphics[width=0.4\textwidth]{I12_r1.pdf}  \hspace{0.5cm}
		   \includegraphics[width=0.5\textwidth]{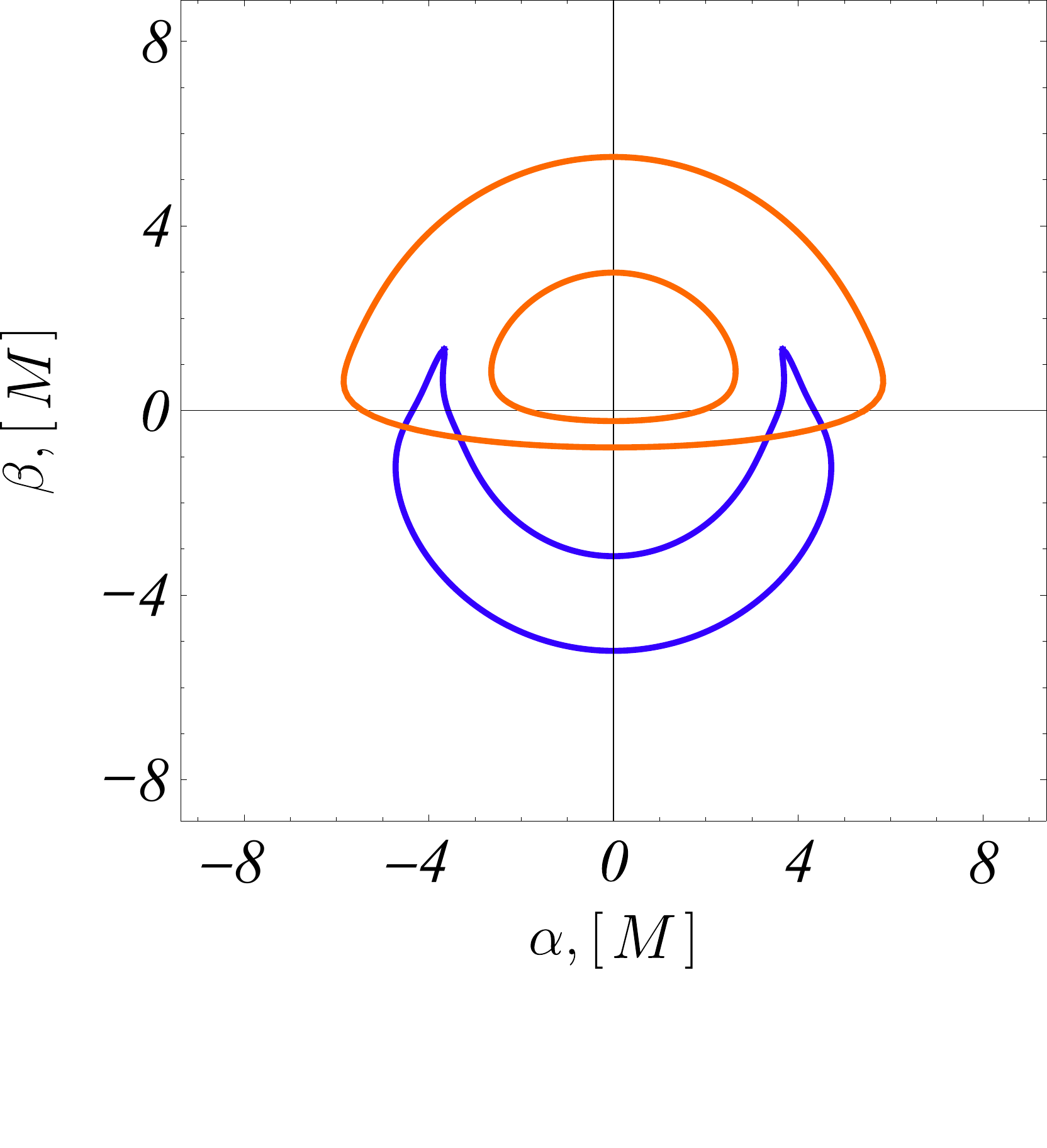} \\[1mm]
           \hspace{0.7cm}  $a)$ \hspace{7.5cm}  $b)$
        \end{tabular}}
 \caption{\label{fig:Image_r1}\small Image formation diagram  for the outer marginally stable orbit $r_{+} = 5.896M$ of the strongly naked singularity with $\gamma=0.48.$ a), and the corresponding observable image b). We use the same conventions as in fig. $\ref{fig:Int_r1}$.}
\end{figure}

Using this intuition we can describe qualitatively the expected image of the outer marginally stable orbit $r_{+}$. We should observe two disconnected direct images ($k=0$), and one secondary image of order $k=1$, which does not close around the origin of the celestial coordinates. In fig. $\ref{fig:Image_r1}$ we present the image of the outer marginally stable orbit computed by the semi-analytical procedure. We can see that it indeed possesses the expected features.

\begin{figure}
    		\setlength{\tabcolsep}{ 0 pt }{\small\tt
		\begin{tabular}{ cc}
           \includegraphics[width=0.4\textwidth]{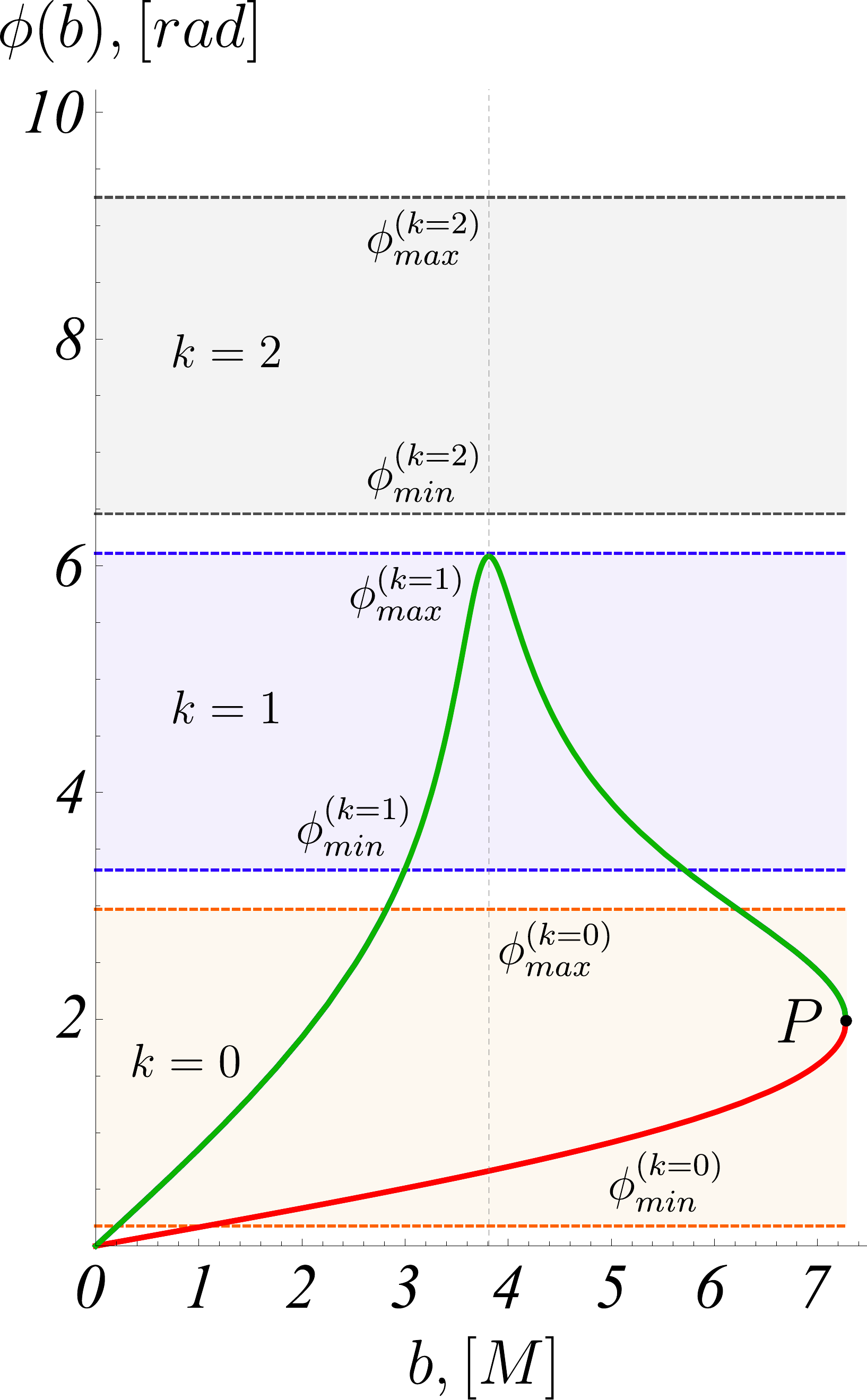}  \hspace{0.5cm}
		   \includegraphics[width=0.5\textwidth]{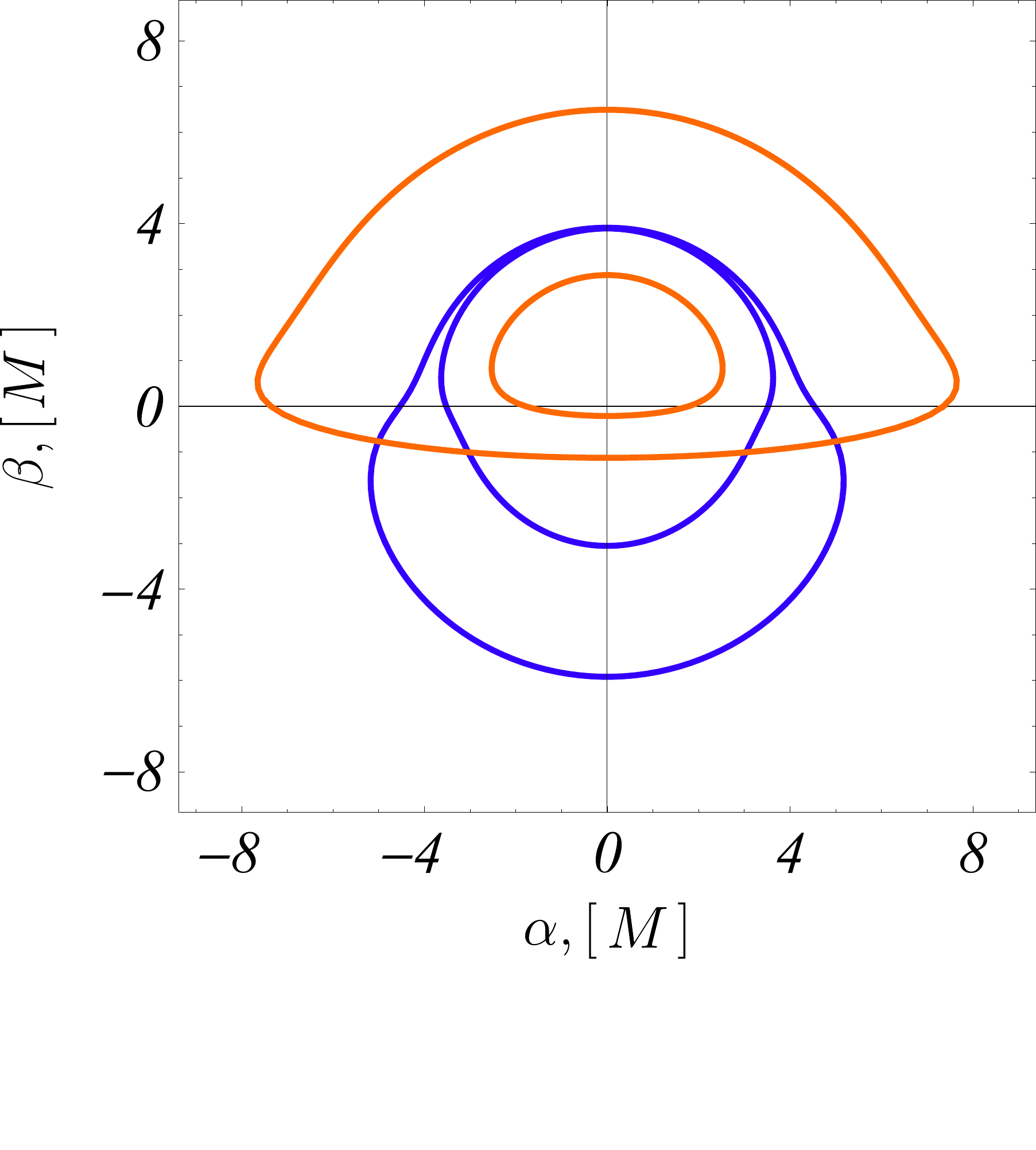} \\[1mm]
           \hspace{0.7cm}  $a)$  \\
           \includegraphics[width=0.4\textwidth]{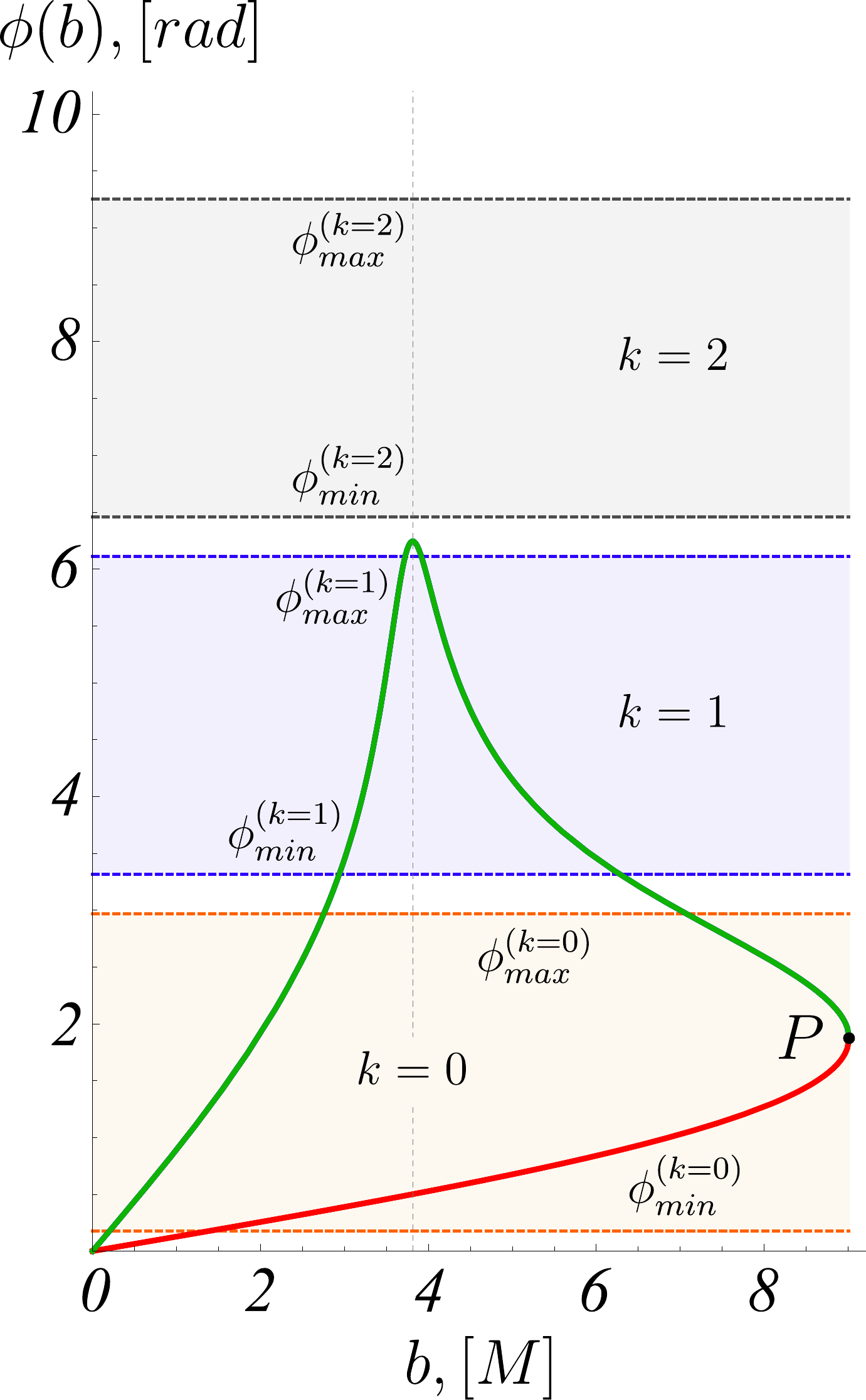}  \hspace{0.5cm}
		   \includegraphics[width=0.5\textwidth]{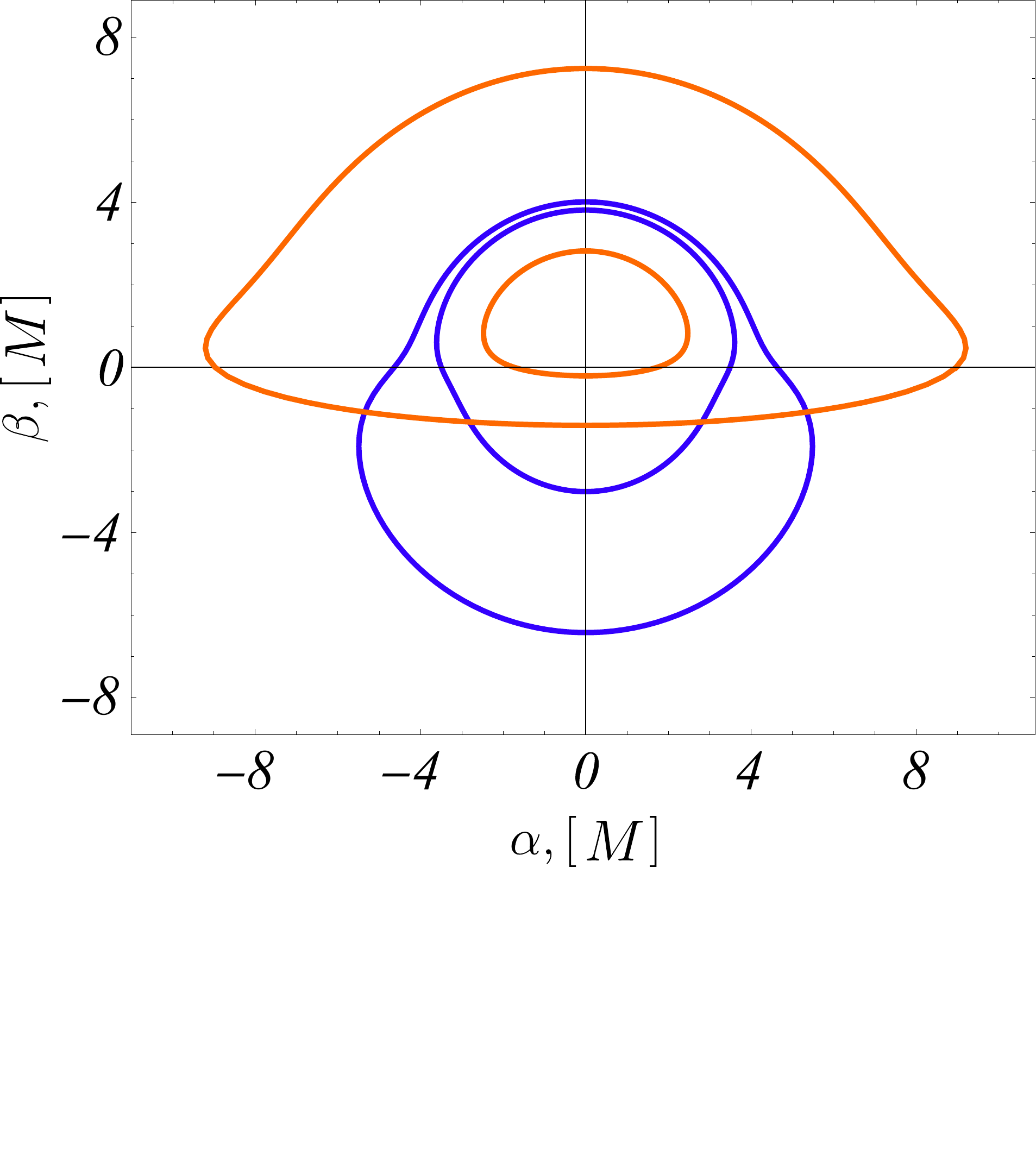} \\[1mm]
           \hspace{0.7cm}  $b)$
                   \end{tabular}}
 \caption{\label{fig:Image_r23}\small Image formation diagrams and the corresponding observable images for the circular orbits located at $r= 7.62M$ a), and $r=9.124M$ b) in the spacetime of the strongly naked singularity with $\gamma=0.48$.}
\end{figure}

\begin{figure}
    		\setlength{\tabcolsep}{ 0 pt }{\small\tt
		\begin{tabular}{ cc}
           \includegraphics[width=0.4\textwidth]{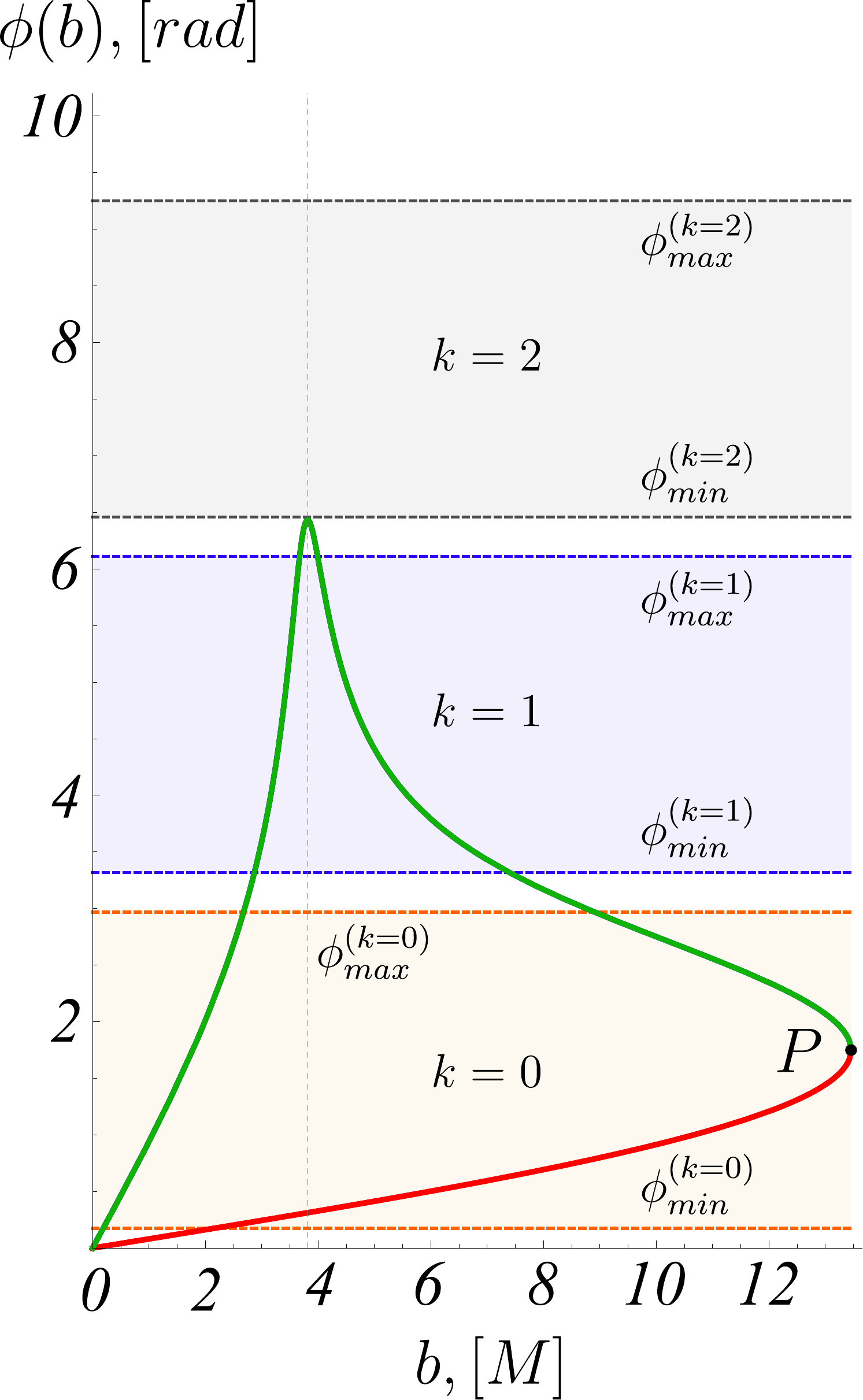}  \hspace{0.5cm}
		   \includegraphics[width=0.5\textwidth]{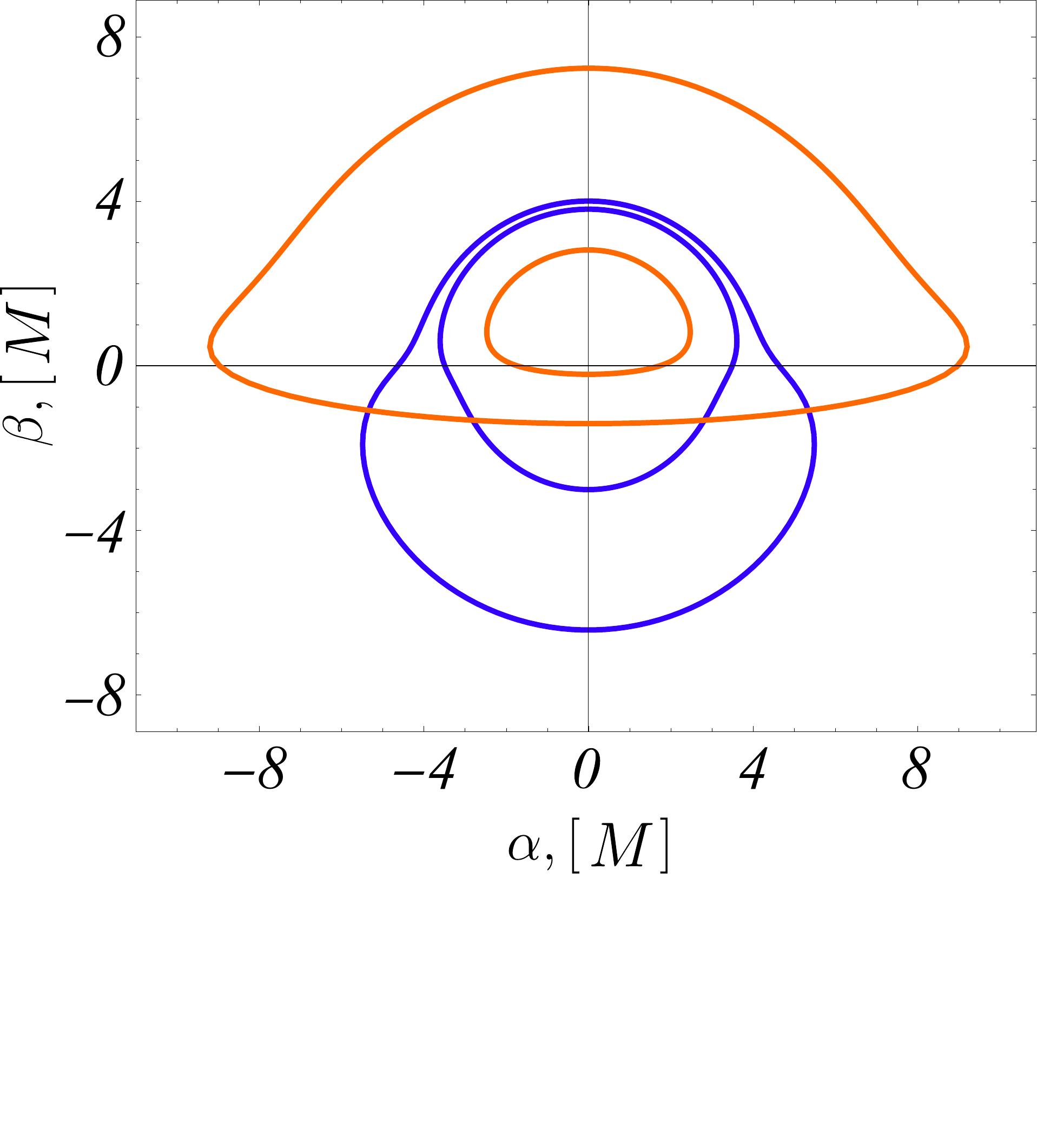} \\[1mm]
           $a)$  \\
          \includegraphics[width=0.4\textwidth]{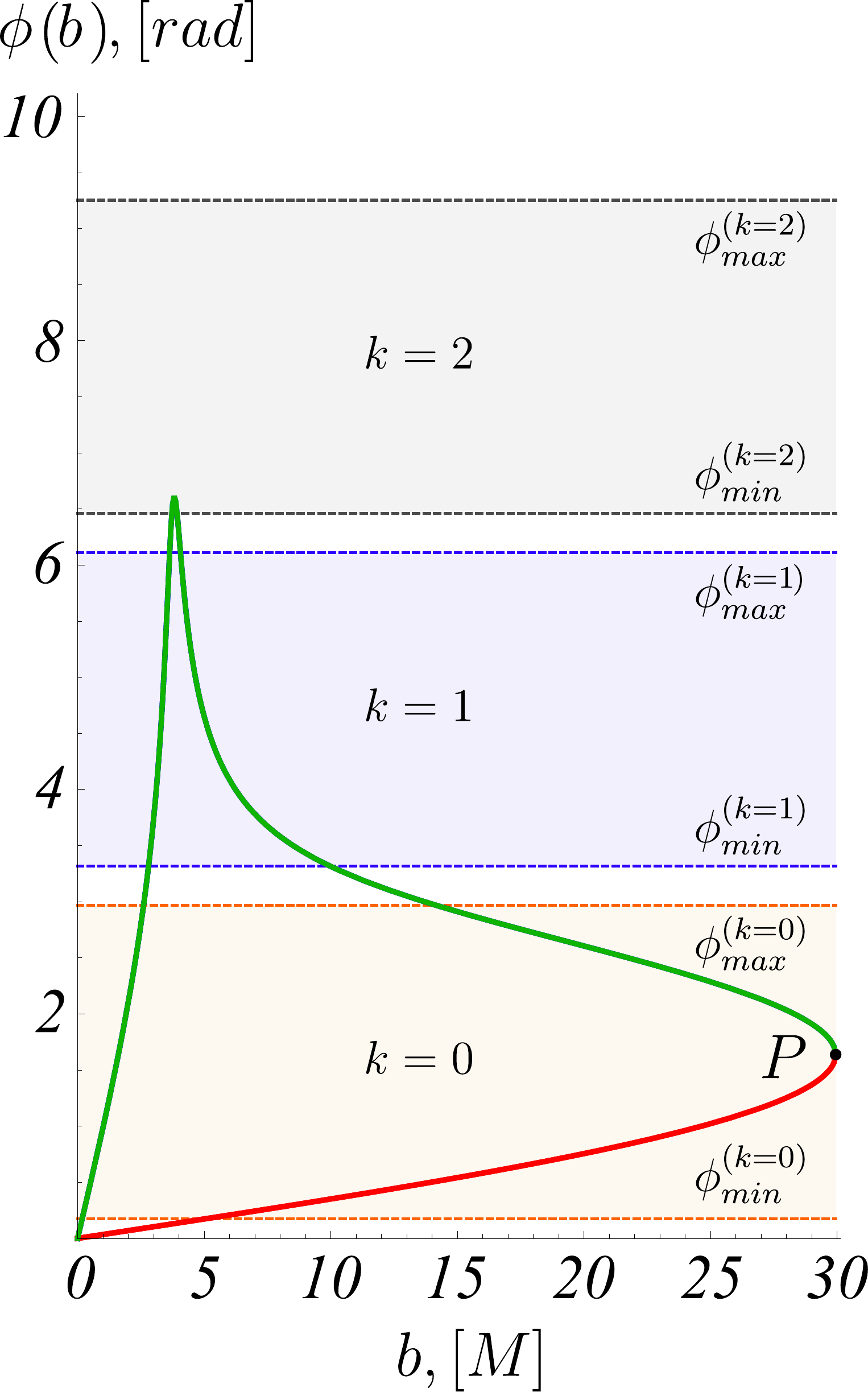}
		   \includegraphics[width=0.6\textwidth]{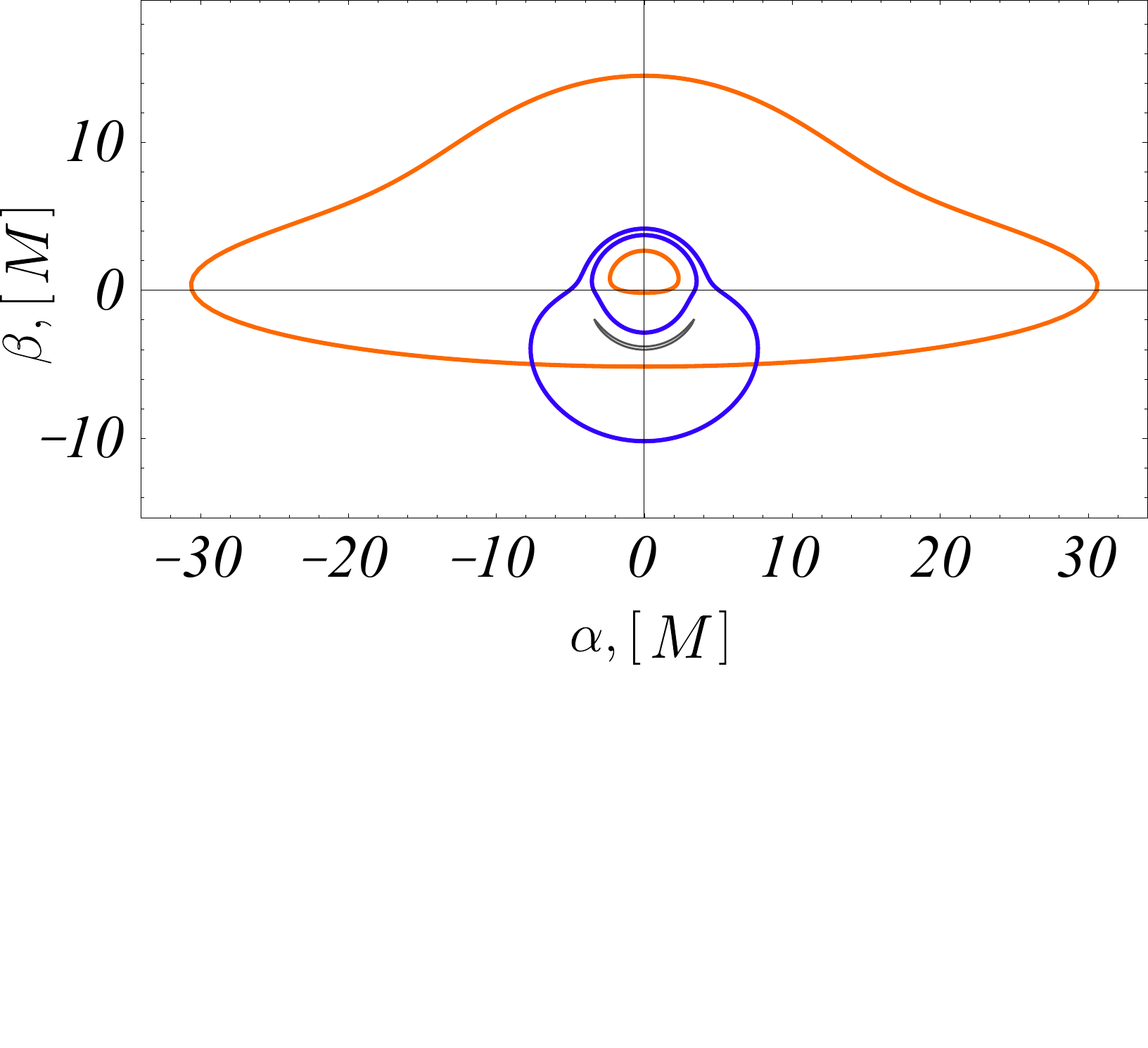} \\[1mm]
          $b)$
                   \end{tabular}}
 \caption{\label{fig:Image_r45}\small Image formation diagrams and the corresponding observable images for the circular orbits located at $r= 13.56M$ a), and $r=30M$ b) in the spacetime of the strongly naked singularity with $\gamma=0.48$.}
\end{figure}

In a similar way, by analyzing the image formation diagrams for the circular orbits with larger radii, we can predict some qualitative features of the optical appearance of the whole outer disk. In figs. $\ref{fig:Image_r23}$-$\ref{fig:Image_r45}$  we present the graphical solutions of eq. ($\ref{orbit_im}$) for several selected circular orbits, each of which marks the appearance of a new feature in the observable image. At the same time the image of each orbit is computed and illustrated for comparison. We see that all the orbits should possess a double image of order $k=0$, which results in the observation of a second copy of the whole outer disk in the form of a central ring. Orbits with radii $r_{+}<r<r_1 \approx 7.62M$ give rise to a secondary image of order $k=1$, which doesn't encircle the origin of the  celestial coordinates. For $r_1 \approx 7.62M$ this secondary image closes, and for $r>r_1$ it decouples into two separate images, which encircle the coordinate origin (see fig. $\ref{fig:Image_r23}$). Thus, circular orbits with radii larger than $r_1$ possess also a double image of order $k=1$. This type of images form another bright ring in the center of the optical appearance of the accretion disk. For orbits with radii $r> r_2\approx 13.56M$ we start to observe a third image of order $k=2$ (see fig. $\ref{fig:Image_r45}$).

\begin{figure}[h!]
\centering
           \includegraphics[width=0.7\textwidth]{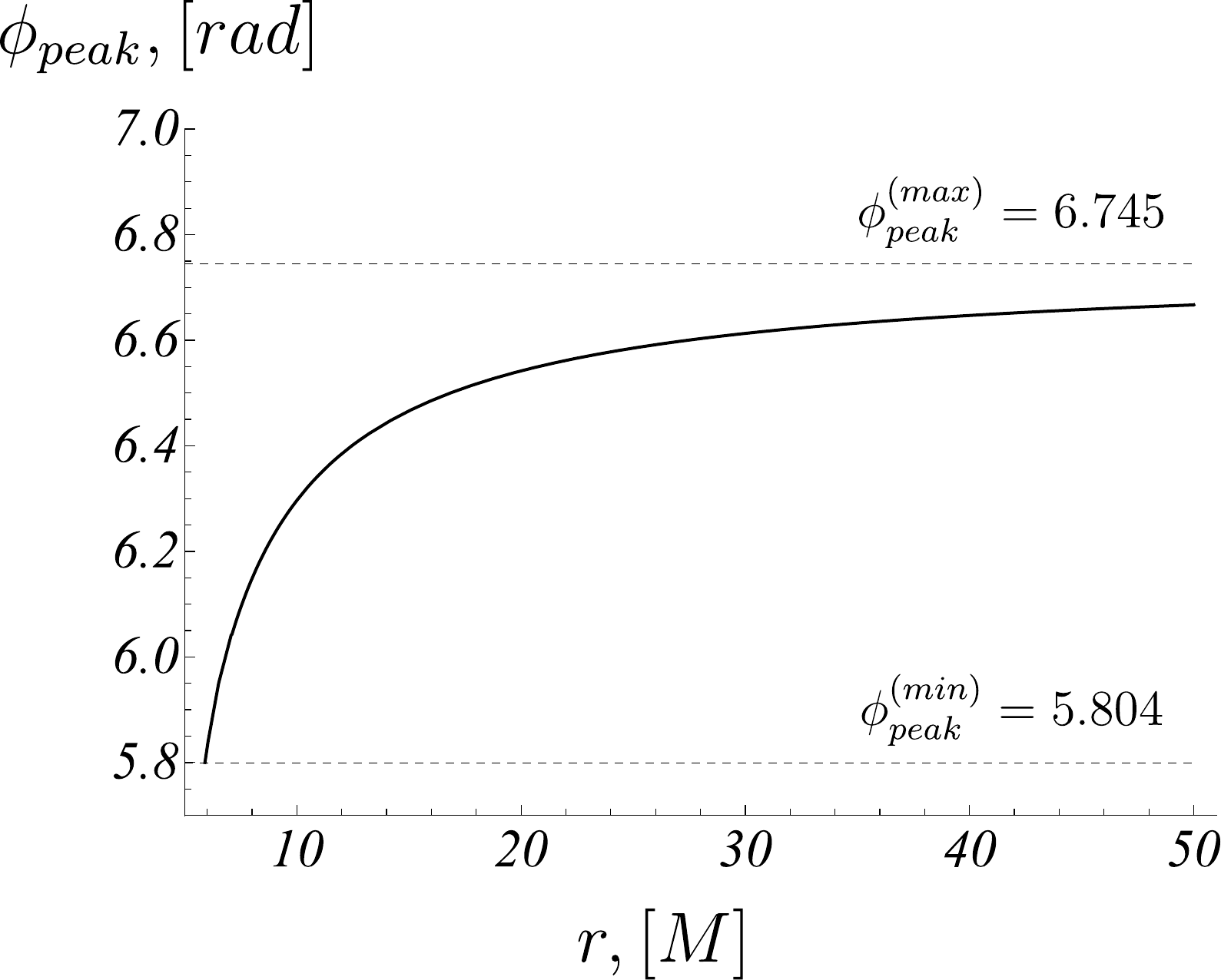}\\[1mm]

 \caption{\label{fig:phi_peak}\small Dependence of the maximum value of the azimuthal angle $\phi_{peak}$ on the radial position of the emitting circular orbit $r_{source}$ for the outer accretion disk. We observe a monotonically increasing function starting at the outer marginally stable orbit $r_{+} = 5.896$  with $\phi_{peak}(r_{+}) = \phi_{peak}^{(min)} = 5.804$ and approaching asymptotically the value $\phi_{peak}^{(max)} = 6.745$. }
\end{figure}

We can prove that $k=2$ is the maximal image order of any circular orbit, which can appear for the chosen solution parameter $\gamma=0.48$ and inclination angle $i= 80^\circ$. The dependence of the azimuthal angle on the impact parameter for the trajectories with a turning point, which is represented by the green curve, is a limited function, and  possesses a single local maximum. This is a distinctive feature of the deflection angle for strongly naked singularities, which was also discussed previously in \cite{Virbhadra:2002}. We can analyze the dependence of the maximal value of the azimuthal angle $\phi_{peak}$ on the radial position of the circular orbit $r_{source}$. Increasing the radius of the orbit the maximum $\phi_{peak}$ monotonically increases approaching asymptotically the value $\phi^{max}_{peak} \approx 6.745$ (see fig. $\ref{fig:phi_peak}$). This value is below the upper limit $\phi^{(k=2)}_{max}$ of the observational window for the images of second order. Consequently, for no circular orbit it will be possible to obtain an intersection with an observation window of higher order.

This property represents a basic distinction from the spacetimes, which possess a photon sphere. In spacetimes with a photon sphere, by definition the integral, which corresponds to the trajectories with turning point, diverges at the value of the impact parameter $b_{ph}$ defining the photon sphere. As a consequence, the green curve tends to infinity when it approaches $b=b_{ph}$, and crosses the observational windows for any order $k$. Thus, images of arbitrary high order  will be observed  for any circular orbit. At the same time the range of the impact parameter, which corresponds to the observable images produced by geodesics with a turning point, shrinks from $b\in[0, b_{max}]$ to $b\in[b_{ph}, b_{max}]$. Then, we cannot have anymore two disconnected portions of the green and the red curves, crossing simultaneously any of the observational windows, and giving rise to double images of certain order. Therefore, the central rings created by this mechanism cannot form. We can conjecture that the formation of double images is a general feature of spacetimes without a photon sphere. It leads to an observable effect visualized by the appearance of the central bright rings, and can be used to distinguish between compact objects without a photon sphere and black holes, for example,  by means of experiments in the electromagnetic spectrum.

In order to illustrate the qualitative behavior for the spacetimes with a photon sphere we present  the image formation diagram for the ISCO of the Schwarzschild solution in fig. $\ref{fig:Int_r1}$ b). We see that the ISCO possesses a single direct image and a single secondary image of order $k=1$, which encircles the coordinate origin, as well as infinitely many higher order images for impact parameters asymptotically approaching that of the photon sphere. These images are, however, observationally undistinguishable from the image of the photon sphere.

\begin{figure}[t!]
    		\setlength{\tabcolsep}{ 0 pt }{\small\tt
		\begin{tabular}{ cc}
           \includegraphics[width=0.33\textwidth]{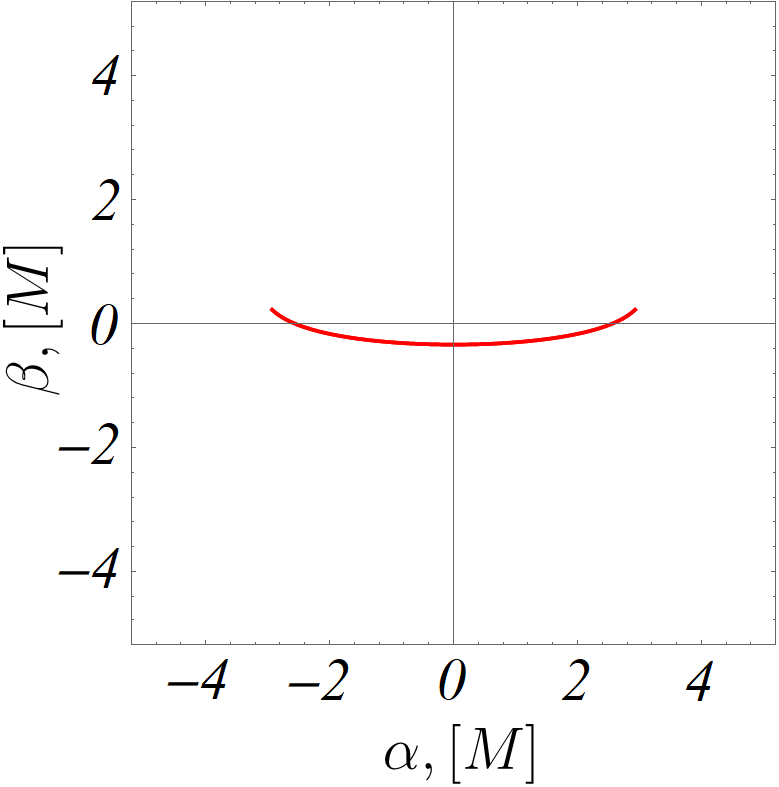}
		   \includegraphics[width=0.33\textwidth]{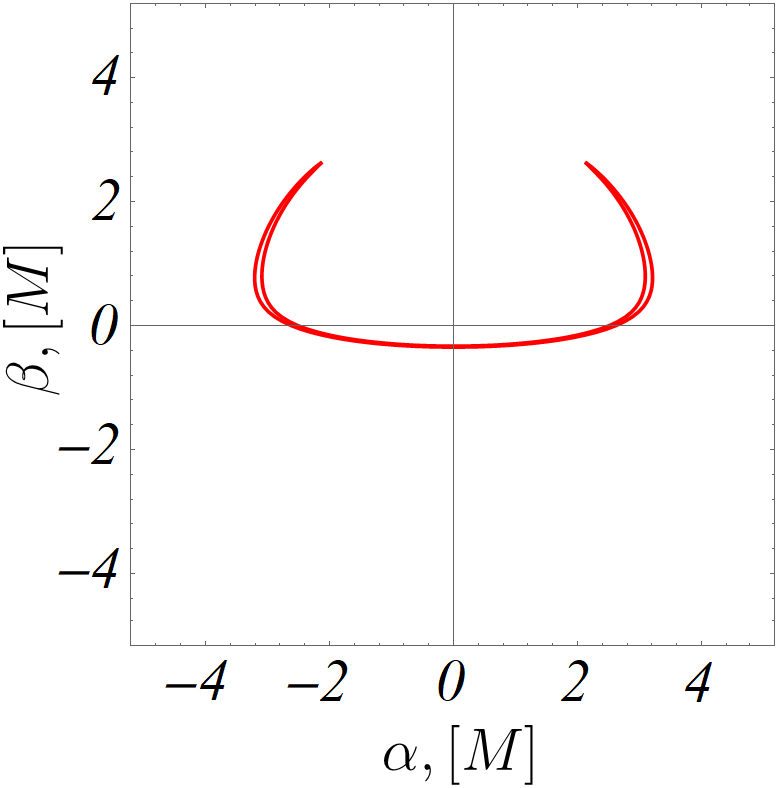}
            \includegraphics[width=0.33\textwidth]{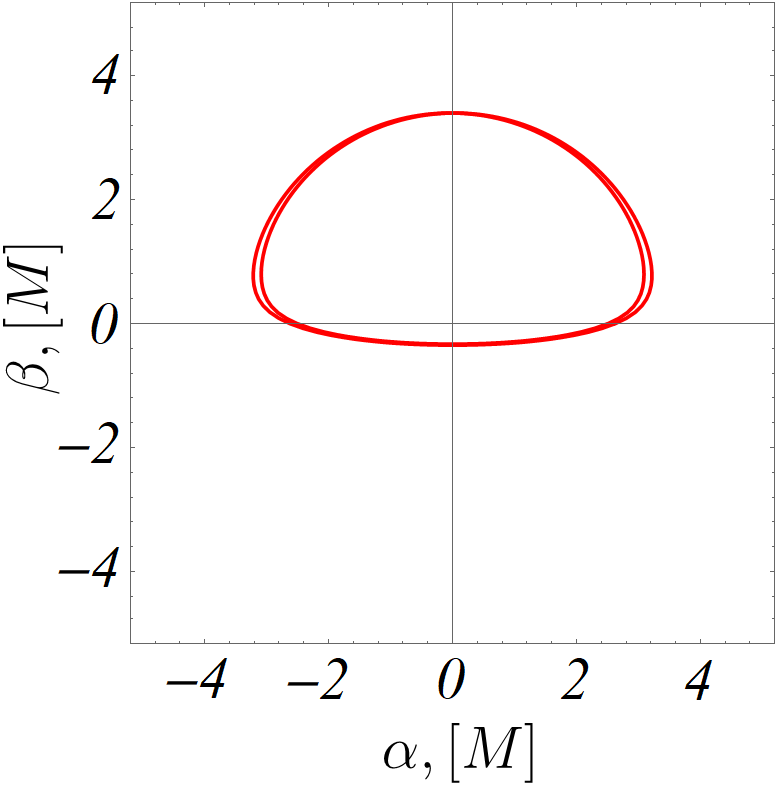} \\[1mm]
           \hspace{0.5cm} $a)\,\,\, r_{source}=4.167M$ \hspace{1cm} $b)\,\,\, r_{source}=4.1681M$ \hspace{1cm} $c) \, \,\, r_{source}=4.1684M$  \\[2mm]
          \includegraphics[width=0.3325\textwidth]{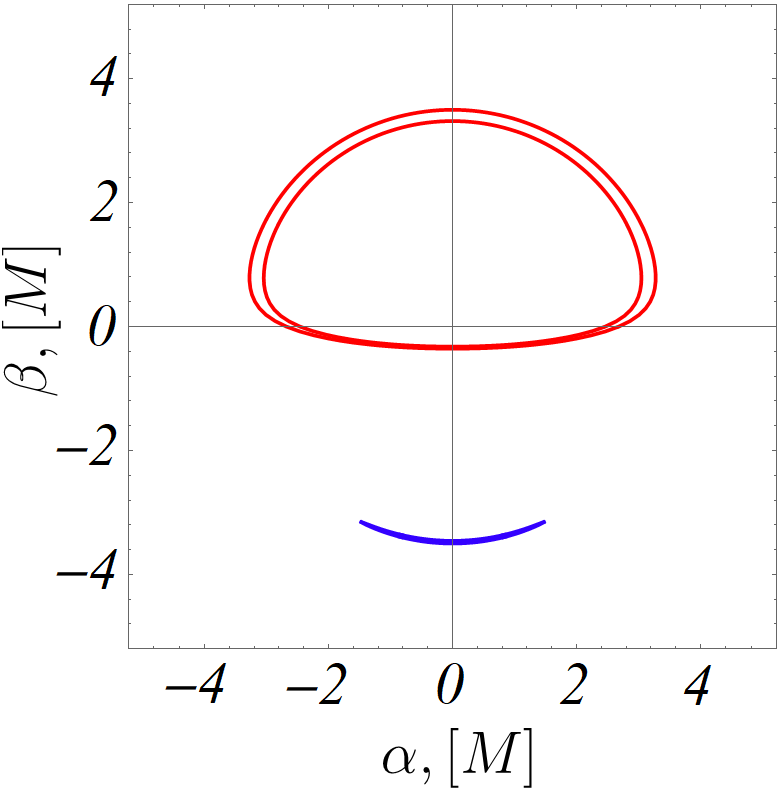}
		   \includegraphics[width=0.3185\textwidth]{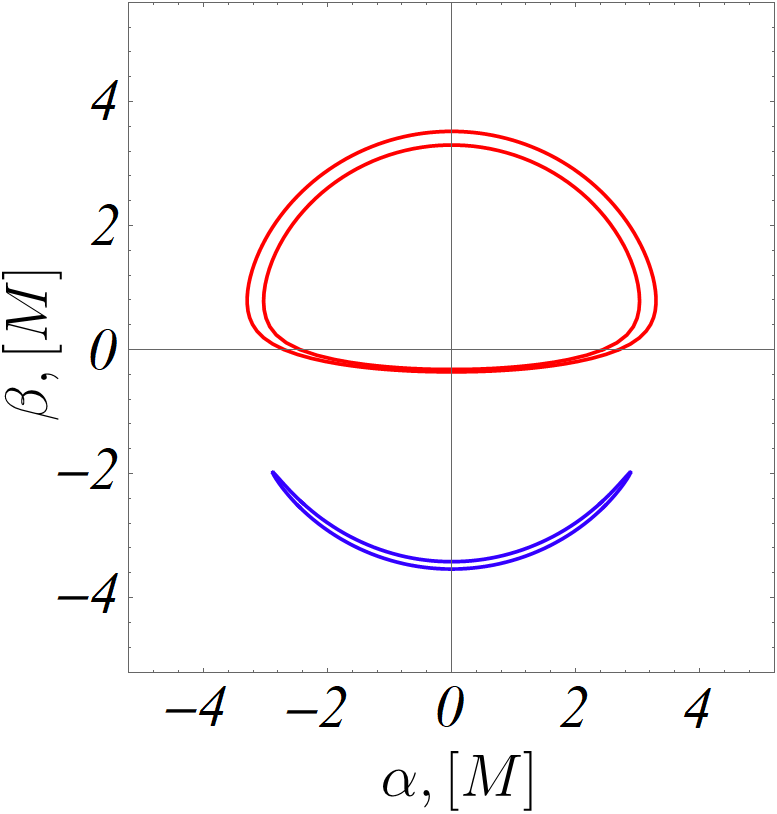}
           \includegraphics[width=0.33\textwidth]{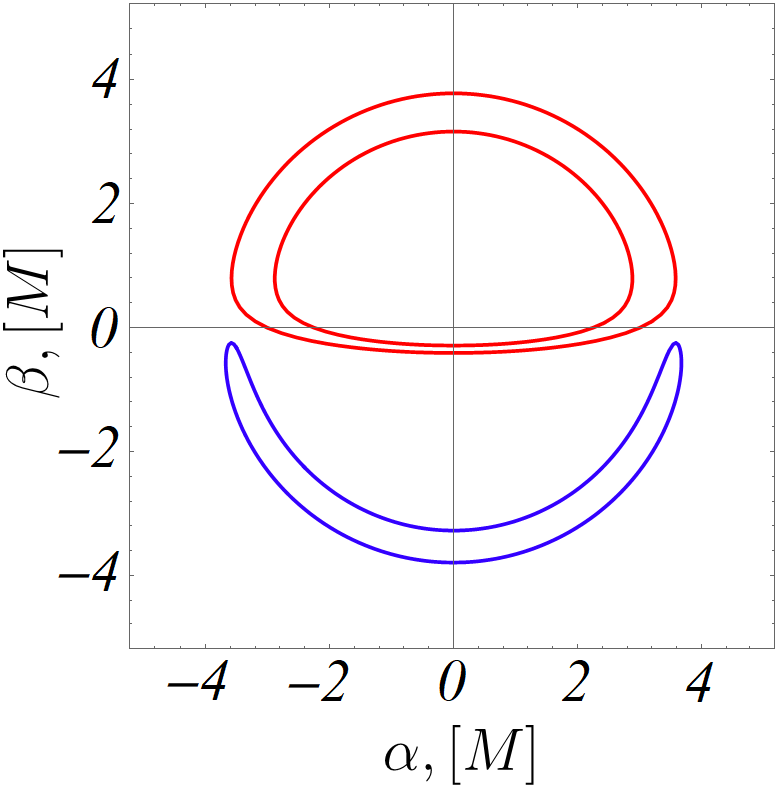} \\[1mm]
          \hspace{0.5cm} $d)\,\,\, r_{source}=4.174M$ \hspace{1cm}  $e)\,\,\, r_{source}=4.177M$ \hspace{1cm} $f)\,\,\, r_{source}=4.271M$
                   \end{tabular}}
 \caption{\label{fig:Image_indisk}\small Images of the circular orbits in the inner accretion disk for the strongly naked singularity with $\gamma=0.48$. The observer is located at $r=5000M$ and at the inclination angle $i=80^\circ$. We illustrate the variety of possible images by considering several successive orbit positions, which are indicated under each plot. We start from an orbit close to the curvature singularity at $r_{source} = 4.167M$ and reach the inner disk limit located at the inner marginally stable orbit $r_{-} =  4.271M$.  }
\end{figure}

Using the same arguments we can analyze the image formation for the circular orbits in the inner accretion disk, which spans between the curvature singularity and the inner marginally stable orbit $r_{-} = 4.27110$. In fig. $\ref{fig:Image_indisk}$ we illustrate the possible qualitatively different images, which can arise for these orbit locations for the strongly naked singularity with $\gamma=0.48$. For small value of the orbit position we observe a single direct image, which does not encircle the coordinate origin (see fig. $\ref{fig:Image_indisk}$ a)). Increasing the radial coordinate, we reach an orbit position $r_1\approx 4.1684M$ when the image closes around the origin (fig. $\ref{fig:Image_indisk}$ c)), and decouples into two disconnected direct images for larger radii (fig.$\ref{fig:Image_indisk}$ d)). At the same time a secondary image of order $k=1$ develops. Going through the remaining circular orbits in the inner disk, we see that the two direct images get more widely separated, while the secondary image grows larger. However, it never closes around the coordinate origin, and it is practically unobservable since it remains hidden behind the direct image of the outer disk (see fig. $\ref{fig:IsoR1}$). The maximal possible image order for the inner disk is $k=1$, since images of higher order cannot be formed for any circular orbit belonging to it. The direct images form a ring, which is observed in the central part of the optical appearance of the full accretion disk (see fig. $\ref{fig:IsoR2}$).

\section{Radiation from the accretion disk}

We consider a physical model for the accretion disk, in which it consists of an anisotropic fluid moving in the equatorial plane with fluid distribution height negligible compared to its extension in the horizontal direction. The model was studied by Novikov and Thorne in \cite{Novikov}, \cite{Page:1974}, where they derived an expression for the flux of the radiant energy over the disk released between its edge $r_i$ and a certain value of the radial coordinate $r$

\begin{equation}\label{F_r}
 F(r)=-\frac{\dot{M}_{0}}{4\pi \sqrt{-g^{(3)}}}\frac{\Omega
_{,r}}{(E-\Omega
L)^{2}}\int_{r_{i}}^{r}(E-\Omega
L)L_{,r}dr.
\end{equation}

\begin{figure}[h!]
\centering
    		\includegraphics[width=0.75\textwidth]{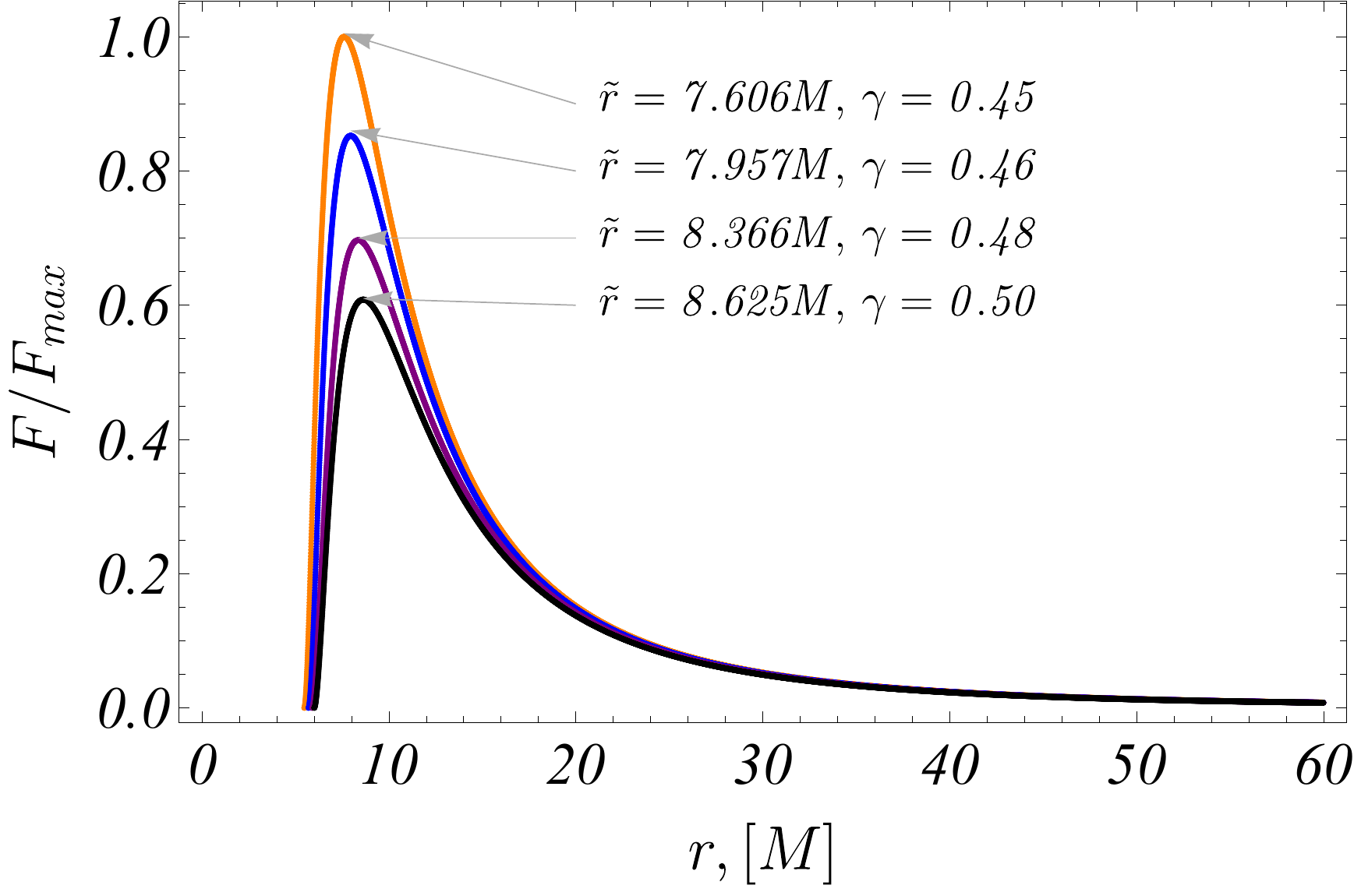}
    \caption{\label{fig:Flux}\small Dependence of the radiation energy flux on the radial coordinate for the outer accretion disk. The curves for several values of the parameter $\gamma$ are depicted in different colour, and the position of the flux maximum $\tilde r$ is specified next to each curve. All the curves are normalized by the maximum of the energy flux for $\gamma=0.45$ $F^{out}_{max} = 4.473\times10^{-6}M\dot{M}$.}
\end{figure}

\begin{figure}[h!]
    		\setlength{\tabcolsep}{ 0 pt }{\small\tt
		\begin{tabular}{ cc}
           \includegraphics[width=0.45\textwidth]{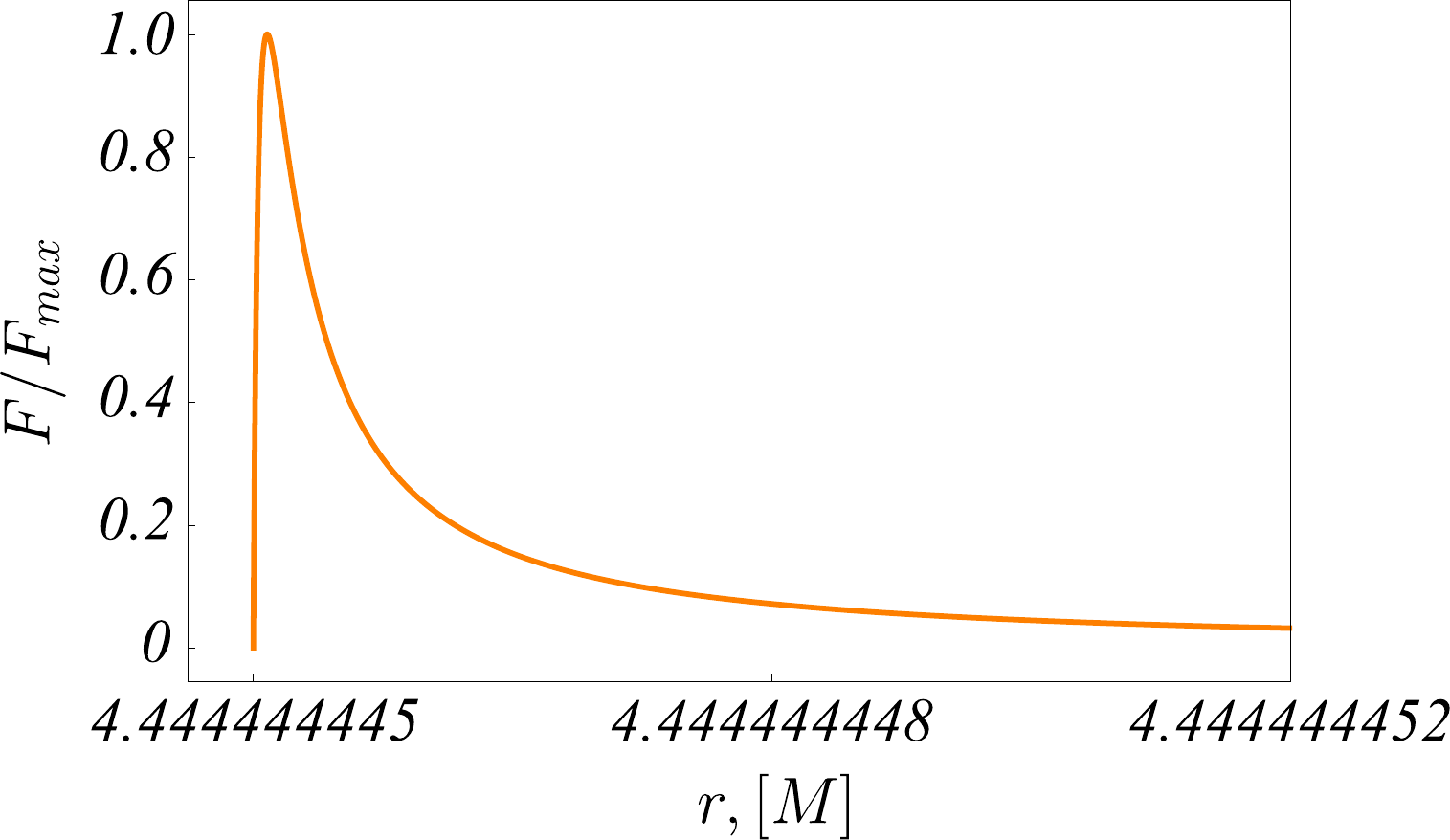} \hspace{0.5cm}
		   \includegraphics[width=0.45\textwidth]{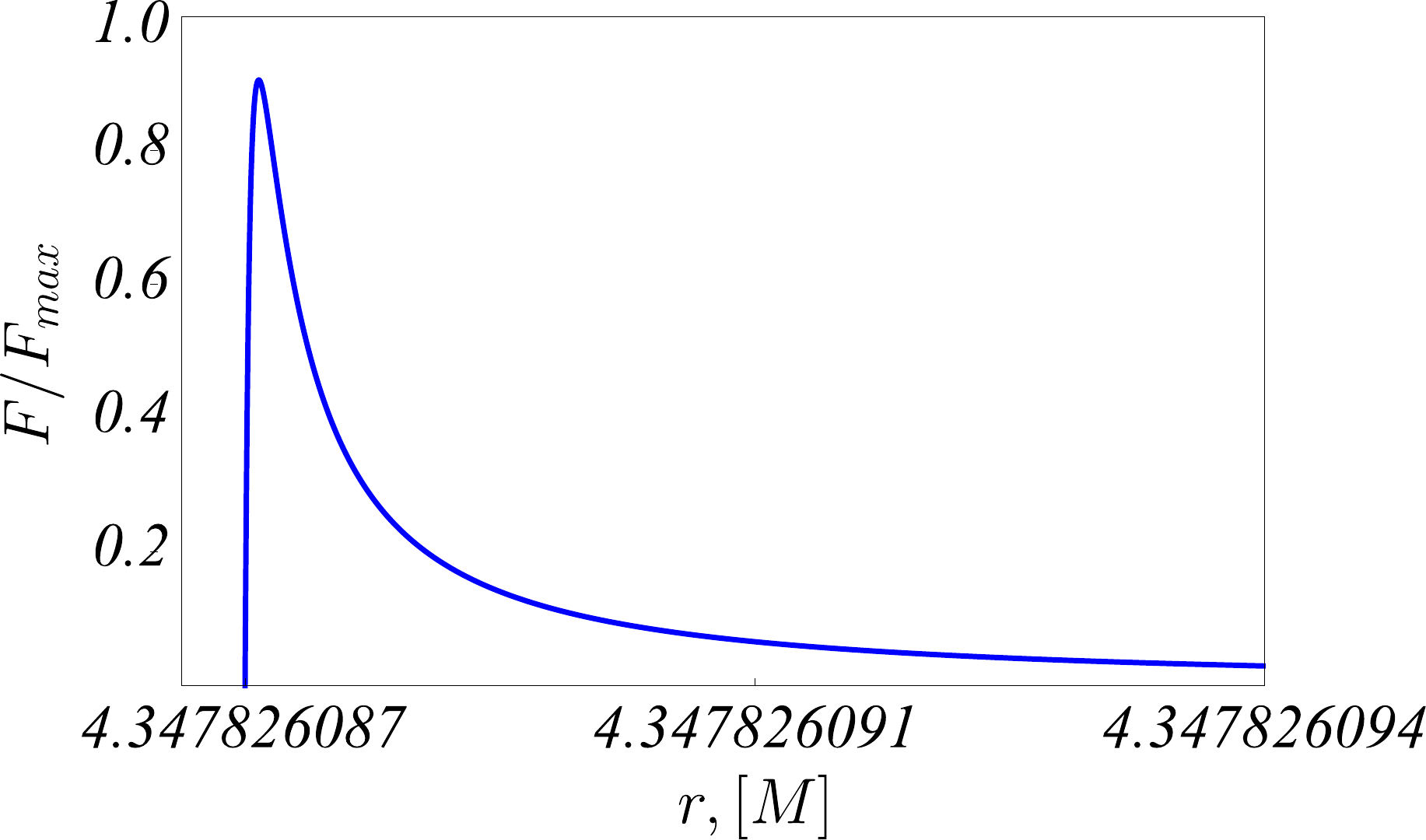} \\[1mm]
          \hspace{0.2cm}$a) \,\,\, \gamma=0.45$ \hspace{5.5cm} $b)\,\,\, \gamma=0.46$ \hspace{3cm}  \\[2mm]
          \includegraphics[width=0.45\textwidth]{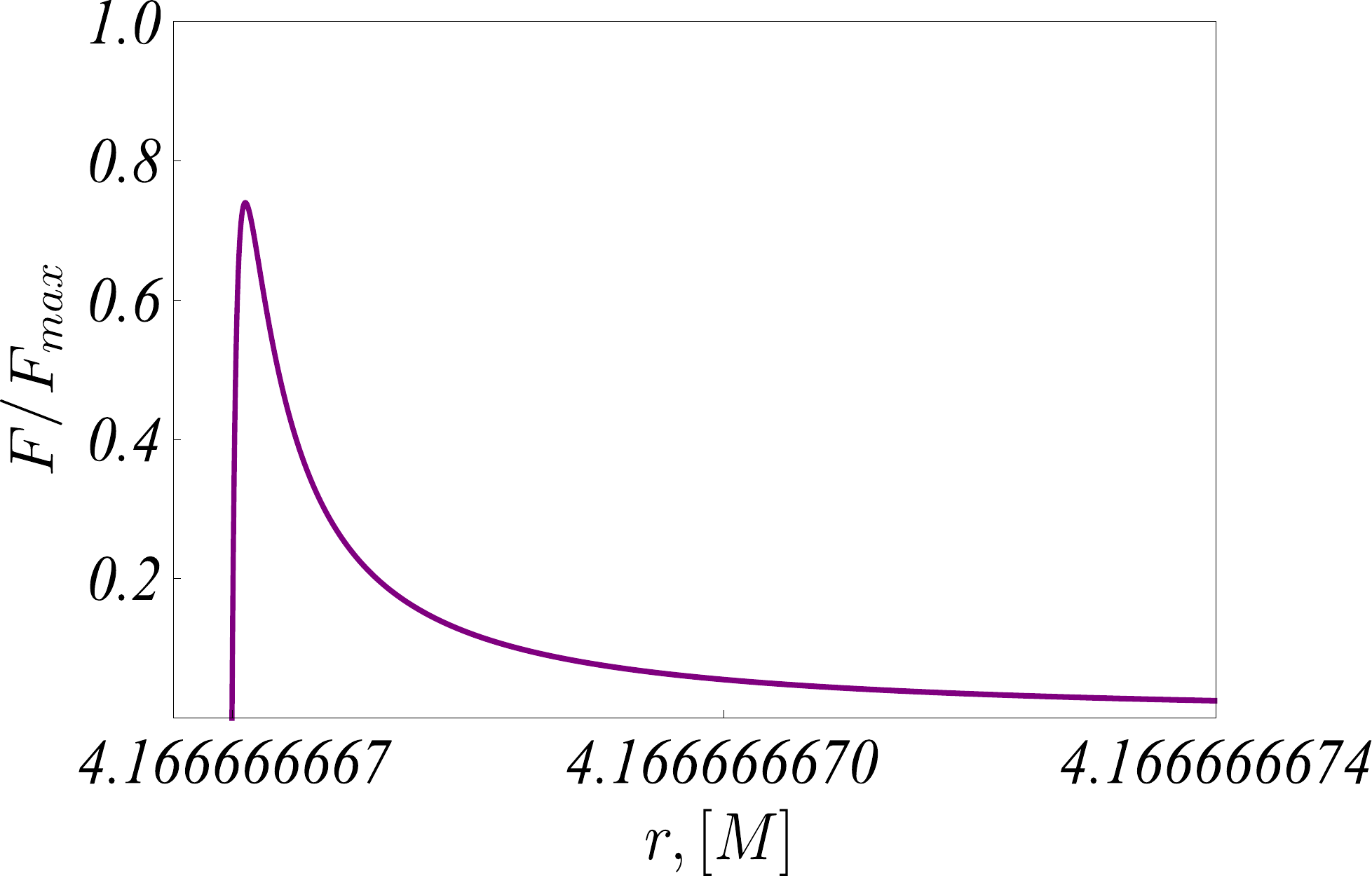}\\[1mm]
          $c) \, \,\, \gamma=0.48$
                   \end{tabular}}
 \caption{\label{fig:FluxIn}\small Dependence of the radiation energy flux on the radial coordinate for the inner accretion disk. All the curves are normalized by the maximum of the energy flux for $\gamma=0.45$ $F^{in}_{max} = 2.275\times10^{9}M\dot{M}$.}
\end{figure}

In the integral we denote by $E$, $L$, and $\Omega$ the specific energy, the specific angular momentum, and the angular velocity of the particles moving on a particular circular orbit, $g^{(3)}$  is the determinant of the induced metric in the equatorial plane, while $\dot M_0$ is a constant representing the mass accretion rate. The kinematic quantities on the orbit can be expressed in the general form

\begin{eqnarray}
E&=&-\frac{g_{tt}}{\sqrt{-g_{tt}-g_{\phi\phi}\Omega^2}},    \label{rotE}  \\
L&=&\frac{g_{\phi\phi}\Omega}{\sqrt{-g_{tt}-g_{\phi\phi}\Omega^2}},     \label{rotL}  \\
\Omega&=&\frac{d\phi}{dt}=\sqrt{-\frac{g_{tt,r}}{g_{\phi\phi,r}}},     \label{rotOmega}
\end{eqnarray}

\noindent
for any static spherically symmetric metric ($\ref{metric_st}$). We compute numerically  the radiation flux for the strongly naked JNW singularity for several values of the solution parameter $\gamma$ in the range $1/\sqrt{5} < \gamma \leq1/2$. The flux consists of two parts emitted by the outer and the inner disk, which we obtain by evaluating the integral with initial conditions $r_i = r_{+}$, and $r_i = r_{cs} + 10^{-10}M$, respectively. We illustrate the radiation flux behavior as a function of the radial coordinate in figs. $\ref{fig:Flux}$-$\ref{fig:FluxIn}$. For each of the two disconnected disk regions the flux possesses a single maximum,  which grows monotonically when the value of $\gamma$ decreases, and its radial position approaches the location of the curvature singularity. In order to be able to compare the cases with different $\gamma$ we normalize all the flux curves by the maximum of the energy flux for $\gamma = 0.45$ for the corresponding disk region, which takes the value $F^{out}_{max} = 4.473\times10^{-6}M\dot{M}$ for the outer disk, and $F^{in}_{max} = 2.275\times10^{9}M\dot{M}$ for the inner disk. We see that the flux emitted by the region close to the curvature singularity is considerably larger in magnitude.


\begin{figure}[h!]
\setlength{\tabcolsep}{ 0 pt }{\footnotesize\tt
		\begin{tabular}{ cc}
           \includegraphics[width=1\textwidth]{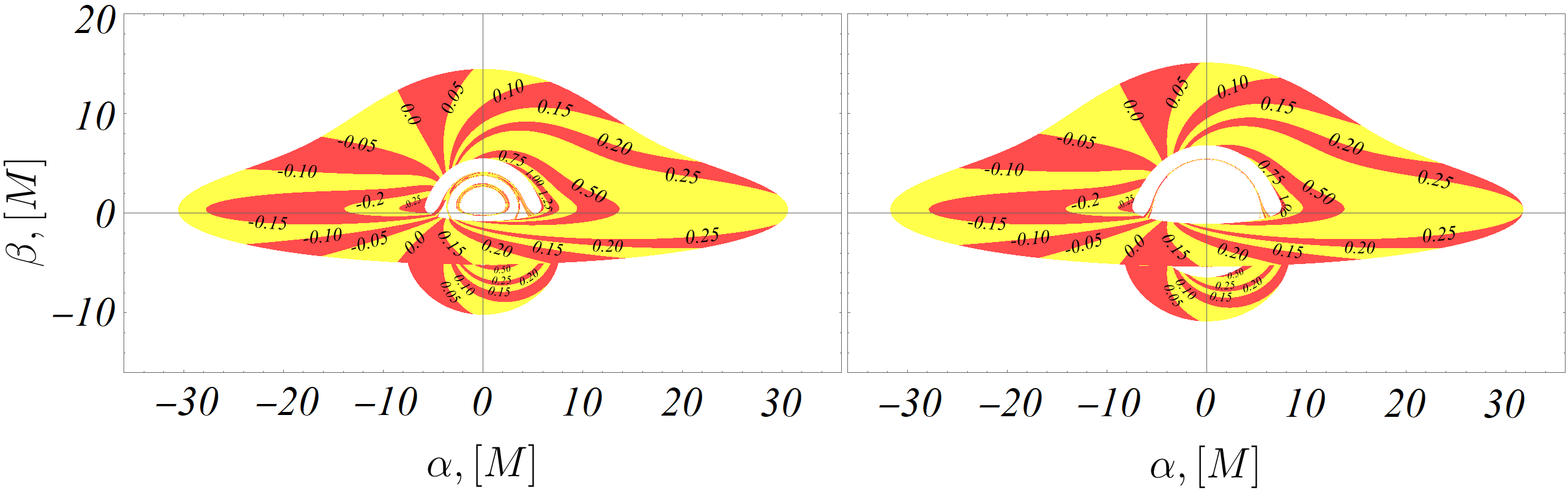}
           \end{tabular}}
 \caption{\label{fig:IsoZ}\small Curves of constant redshift $z$ for the outer disk of the strongly naked singularity with $\gamma=0.48$ (left), and the Schwarzschild black hole (right). The observer is located at $r_{obs} = 5000M$, and at the inclination angle $i=80^\circ$. The value of the relevant redshift is specified on each contour.}
\end{figure}

\begin{figure}[h!]
\centering
\setlength{\tabcolsep}{ 0 pt }{\footnotesize\tt
		\begin{tabular}{ c}
           \includegraphics[width=0.5\textwidth]{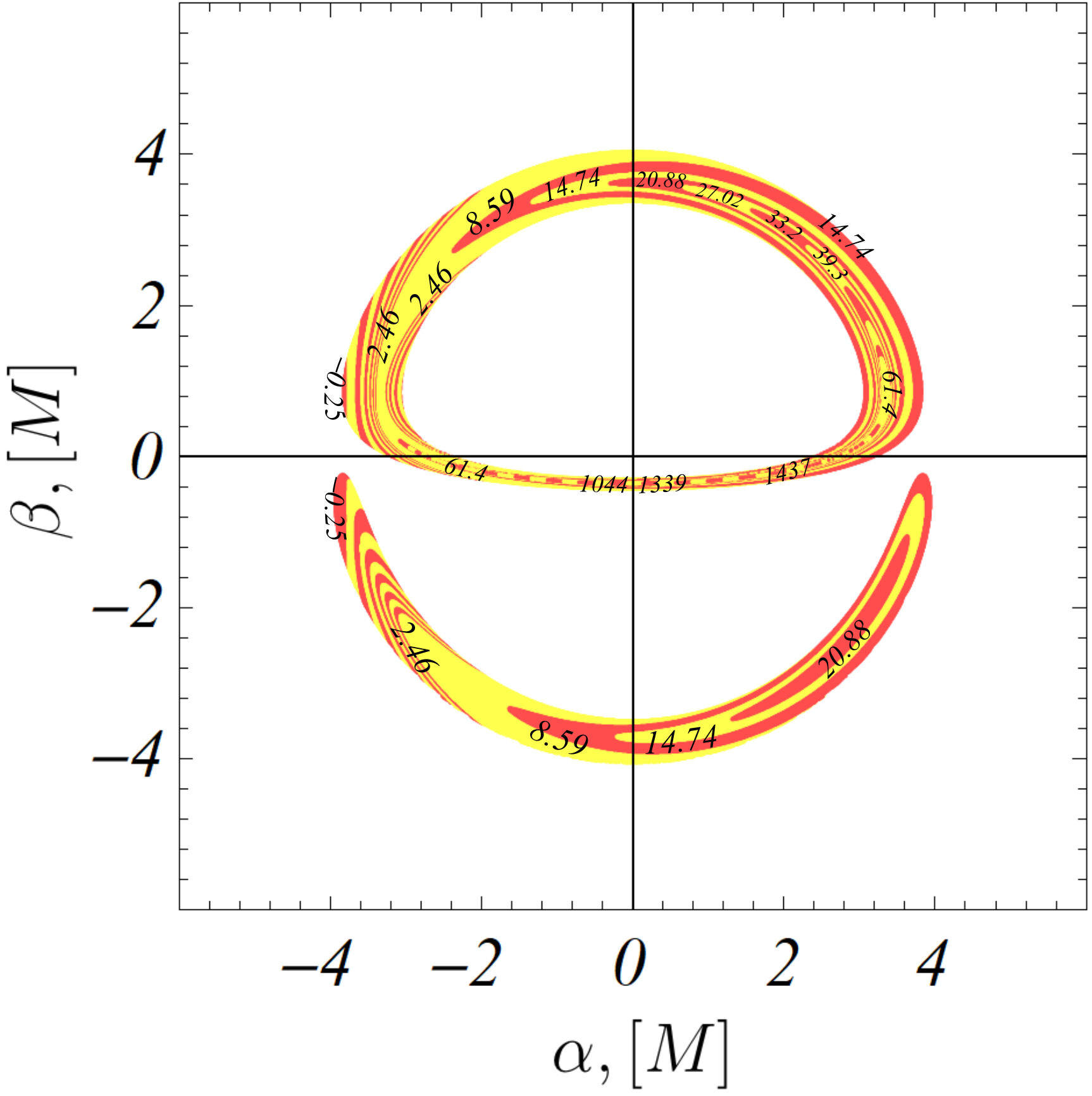}
         \end{tabular}}
 \caption{\label{fig:IsoZIn}\small Curves of constant redshift $z$  for the inner disk of the strongly naked singularity with $\gamma=0.48$. The observer is located at $r_{obs} = 5000M$, and at the inclination angle $i=80^\circ$.}
\end{figure}

In order to obtain the observable radiation flux for a distant observer, we should consider its modification by the gravitational redshift. The apparent intensity $F_{obs}$ in each point of the observer's sky is given by

\begin{equation}\label{F_obs}
F_{obs} = \frac{F}{(1+z)^4},
\end{equation}
where $z$ is the gravitational redshift. For a general static spherically symmetric metric it can be expressed in the form \cite{Luminet:1979}

\begin{equation}
1+z=\frac{1+\Omega b}{\sqrt{ -g_{tt} - \Omega^2 g_{\phi\phi}}},
\end{equation}
where $b=L/E$ is the impact parameter related to the celestial coordinates $\alpha$ and $\beta$ by means of eqs. ($\ref{p_alpha}$). We study the variation of the gravitational redshift in fig. $\ref{fig:IsoZ}$, where we present its isolines for the strongly naked Janis-Newman-Winicour singularity with $\gamma = 0.48$, and the Schwarzschild black hole for comparison. In both cases we obtain very similar redshift distributions, observing a region of negative redshift for negative values of the celestial coordinate $\alpha$, which will result in the increase of the values of the observable radiation flux in the area. In order to make the image more feasible  we illustrate in  fig. $\ref{fig:IsoZ}$ only the redshift for the outer disk of the strongly naked singularity, and present its variation for the inner disk separately in fig. $\ref{fig:IsoZIn}$.

\begin{figure}[h!]
\setlength{\tabcolsep}{ 0 pt }{\footnotesize\tt
		\begin{tabular}{ cc}
           \includegraphics[width=1\textwidth]{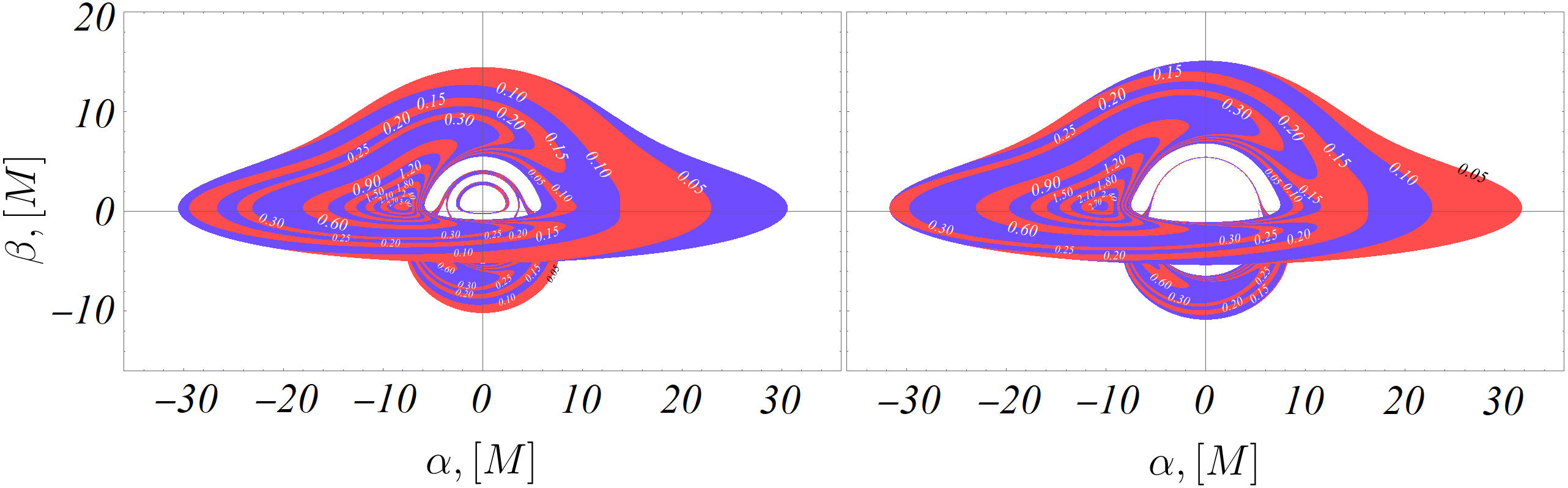}
		\end{tabular}}
 \caption{\label{fig:IsoFlux}\small Contour plot of the apparent radiation flux for the outer disk of the strongly naked singularity with $\gamma=0.48$ (left) and the Schwarzschild black hole (right). The observer is located at $r_{obs} = 5000M$, and at the inclination angle $i=80^\circ$. The observable flux is normalized by the maximal value of the radial flux distribution for each solution. }
\end{figure}

\begin{figure}[h!]
\setlength{\tabcolsep}{ 0 pt }{\footnotesize\tt
		\begin{tabular}{ cc}
           \includegraphics[width=1\textwidth]{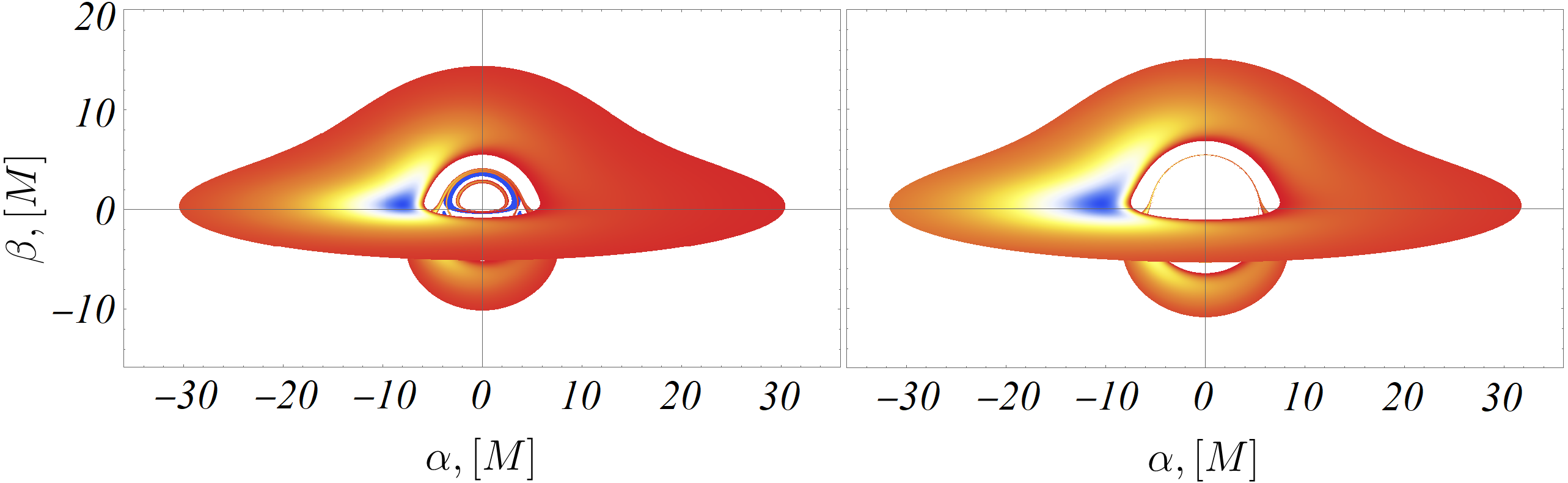}
		\end{tabular}}
 \caption{\label{fig:ColorDisk}\small Continuous distribution of the apparent radiation flux for the strongly naked singularity with $\gamma=0.48$ (left) and the Schwarzschild black hole (right). The observer is located at $r_{obs} = 5000M$, and at the inclination angle $i=80^\circ$. }
\end{figure}

\begin{figure}[h!]
\setlength{\tabcolsep}{ 0 pt }{\footnotesize\tt
		\begin{tabular}{ cc}
           \includegraphics[width=0.45\textwidth]{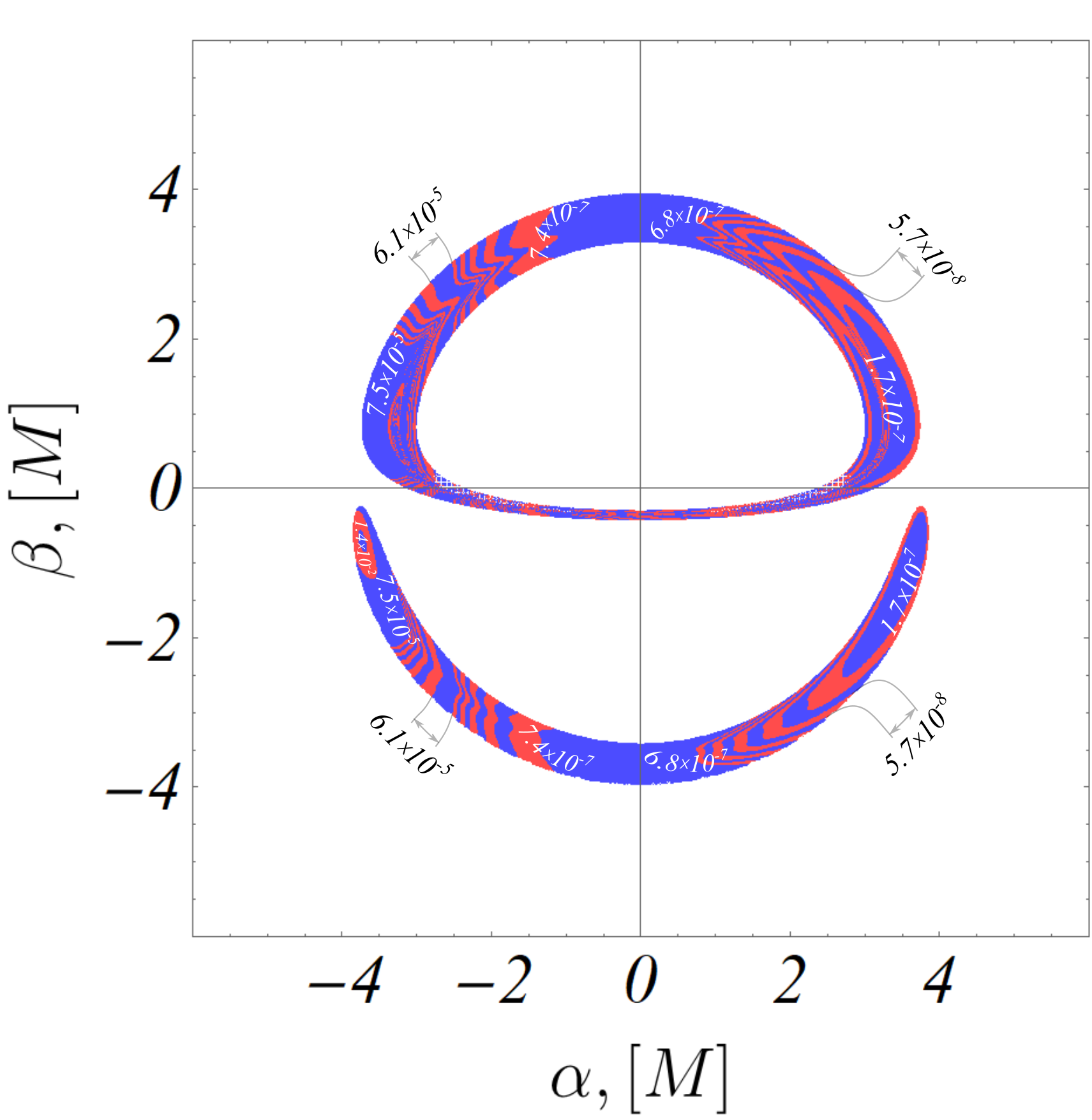}\hspace{0.5cm}
		   \includegraphics[width=0.445\textwidth]{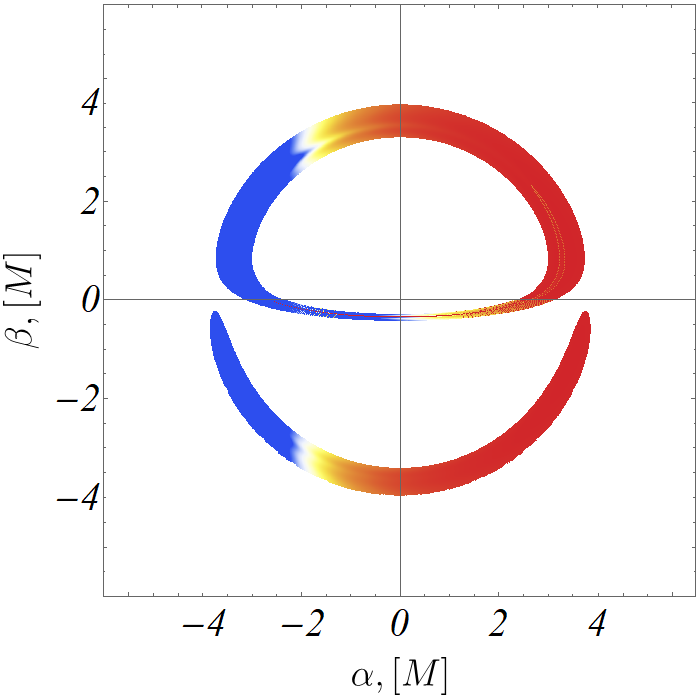}
        \end{tabular}}
 \caption{\label{fig:ColorDiskIn}\small Curves of constant observable flux (left) and its continuous distribution (right) for the inner disk of the strongly naked singularity with $\gamma=0.48$.}
\end{figure}

In figs. $\ref{fig:IsoFlux}$ -$\ref{fig:ColorDiskIn}$ we investigate the apparent flux intensity $F_{obs}$, given by eq. ($\ref{F_obs}$) by constructing two types of images for the strongly naked singularity with $\gamma = 0.48$. For comparison we also provide the flux distribution for the Schwarzschild solution. On the one hand we present contour plots, giving the flux isolines, where the apparent flux is normalized by the maximum value of the radial flux distribution ($\ref{F_r}$) for the particular solution. As for the redshift distribution, in order to make  the image more clear we illustrate only the outer disk in the contour plots, and provide a separate figure for the inner disk (see fig. $\ref{fig:ColorDiskIn}$).  On the other hand we illustrate the distribution of the apparent radiant energy by means of color images. Each flux intensity is visualized in a certain colour of the spectrum from red to blue by means of a continuous color map, as red is associated with the lowest intensity, while blue corresponds to the maximal values.

In the outer disk image the flux intensity for the strongly naked singularity has a similar distribution as for the Schwarzschild solution. For both solutions there is a local flux maximum in the region of negative redshift near the apparent position of the orbit with the maximum radiation energy ($\ref{F_r}$). For the strongly naked singularity the peak reaches a higher value, and the apparent emitted radiation is more concentrated around its location, while for the Schwarzschild solution the flux is more evenly distributed, and grows more gradually towards the maximum. The global maximum of the radiation intensity for the strongly naked singularity is observed at the location of the central ring corresponding to the image of the inner disk, which is represented in dark blue in the color images. The observable flux intensity from the inner disk is $10^{10}$ times larger than the radiation from the outer disk. We can give as an estimate of the flux magnitudes the ratio of the maximum values for the inner and outer disk $F^{in}_{obs}/ F^{out}_{obs} = 2.49\times10^{10}$. For this reason, in order to be able to visualize the distribution of the observable flux from the outer disk,  all the flux values in the color images in Fig. $\ref{fig:ColorDisk}$ are normalized by the maximum of the observable radiation intensity for the outer disk. Unfortunately, in this way we cannot illustrate the variation of the flux from the inner disk due to the large difference in the magnitudes of the fluxes. Therefore, we provide in Fig. $\ref{fig:ColorDiskIn}$  a separate color image for the observable radiation of the inner disk, in which the flux is normalized by its own maximal value.

\section{Conclusion}

We study the optical appearance of  a geometrically thin and optically thick accretion disk around the strongly naked Janis-Newman-Winicour singularity. The solution represents an example of a compact object spacetime, which possesses no photon sphere. As such it provides intuition about the observable effects, which we can expect in this physical situation. Another specific feature of the strongly naked singularity is that the stable circular orbits in the equatorial plane are distributed in two disconnected regions. As a result, the thin accretion disk possesses a characteristic structure. It consists of two separate portions - an inner disk spanning from the curvature singularity to the inner marginally stable orbit, and an outer disk extending from the outer marginally stable orbit to infinity.

The absence of a photon sphere and the particular structure of the accretion disk lead to the formation of various observable images, which don't exist in the case of the Schwarzschild black hole. The possible order of the images for the strongly naked singularity is restricted by $k=2$ and the outer accretion disk possesses double direct image and a double secondary image of order $k=1$. Thus, we observe a second copy of the outer disk in the central region of the image in the form of two bright rings. In the case of spacetimes with a photon sphere this region is dark, and corresponds to the compact object's shadow. The inner disk is visualized as a third central ring with a slightly larger radius than the image of the outer disk.

We explain in detail the physical reasons for the formation of the different images, connecting them with certain properties of the photon trajectories in the strongly naked Janis-Newman-Winicour spacetime. We further evaluate the observable radiation intensity from the accretion disk using the Novikov-Thorne model. The radiation from the central rings reaches  the maximal observable intensity within the whole accretion disk, so it should be detectable in the experiments. Thus, the image of the accretion disk can be used  to  distinguish black holes from strongly naked singularities in the electromagnetic observations.  We make the conjecture that a similar ring structure will also be present in other spacetimes without a photon sphere, and it can be used as a general signature for differentiating observationally between black holes and compact objects possessing no photon sphere.

\section*{Acknowledgments}
We gratefully acknowledge support by the Bulgarian NSF Grant KP-06-H38/2. P.N. is partially supported by the Bulgarian NSF Grant DM 18/3. J.K. and P.N. acknowledge support by the DFG Research Training Group 1620 “Models of Gravity”. S. Y. acknowledges financial support by the Bulgarian NSF Grant No. KP-06-H28/7. Networking support by the COST Actions CA16104 and CA16214 is also gratefully acknowledged.

\end{document}